\documentclass[twocolumn]{aastex631}

\usepackage{amsmath} 
\usepackage{hyperref}

\begin{document}

\title{The Importance of Dust Distribution in Ionizing-photon Escape: \\NIRCam and MIRI Imaging of a Lyman Continuum-emitting Galaxy at $z\sim3.8$}

\correspondingauthor{Zhiyuan Ji}
\email{zhiyuanji@arizona.edu}

\author[0000-0001-7673-2257]{Zhiyuan Ji}
\affiliation{Steward Observatory, University of Arizona, 933 N. Cherry Avenue, Tucson, AZ 85721, USA}

\author[0000-0002-8909-8782]{Stacey Alberts}
\affiliation{AURA for the European Space Agency (ESA), Space Telescope Science Institute, 3700 San Martin Drive, Baltimore, MD 21218, USA}
\affiliation{Steward Observatory, University of Arizona, 933 N. Cherry Avenue, Tucson, AZ 85721, USA}

\author[0000-0003-3307-7525]{Yongda Zhu}
\affiliation{Steward Observatory, University of Arizona, 933 N. Cherry Avenue, Tucson, AZ 85721, USA}

\author[0000-0002-5057-135X]{Eros Vanzella}
\affiliation{INAF – OAS, Osservatorio di Astrofisica e Scienza dello Spazio di Bologna, Via Gobetti 93/3, I-40129 Bologna, Italy}

\author[0000-0002-7831-8751]{Mauro Giavalisco}
\affiliation{University of Massachusetts Amherst, 710 North Pleasant Street, Amherst, MA 01003-9305, USA}

\author[0000-0003-4565-8239]{Kevin Hainline}
\affiliation{Steward Observatory, University of Arizona, 933 N. Cherry Avenue, Tucson, AZ 85721, USA}

\author[0000-0003-0215-1104]{William M. Baker}
\affiliation{DARK, Niels Bohr Institute, University of Copenhagen, Jagtvej 128, DK-2200 Copenhagen, Denmark}

\author[0000-0002-8651-9879]{Andrew J.\ Bunker}
\affiliation{Department of Physics, University of Oxford, Denys Wilkinson Building, Keble Road, Oxford OX1 3RH, UK}

\author[0000-0003-4337-6211]{Jakob M. Helton}
\affiliation{Steward Observatory, University of Arizona, 933 N. Cherry Avenue, Tucson, AZ 85721, USA}

\author[0000-0002-6221-1829]{Jianwei Lyu}
\affiliation{Steward Observatory, University of Arizona, 933 N. Cherry Avenue, Tucson, AZ 85721, USA}

\author[0000-0002-5104-8245]{Pierluigi Rinaldi}
\affiliation{Steward Observatory, University of Arizona, 933 N. Cherry Avenue, Tucson, AZ 85721, USA}

\author[0000-0002-4271-0364]{Brant Robertson}
\affiliation{Department of Astronomy and Astrophysics University of California, Santa Cruz, 1156 High Street, Santa Cruz CA 96054, USA}

\author[0000-0003-4770-7516]{Charlotte Simmonds}
\affiliation{Kavli Institute for Cosmology, University of Cambridge, Madingley Road, Cambridge, CB3 0HA, UK}
\affiliation{Cavendish Laboratory, University of Cambridge, 19 JJ Thomson Avenue, Cambridge, CB3 0HE, UK}

\author[0000-0002-8224-4505]{Sandro Tacchella}
\affiliation{Kavli Institute for Cosmology, University of Cambridge, Madingley Road, Cambridge, CB3 0HA, UK}
\affiliation{Cavendish Laboratory, University of Cambridge, 19 JJ Thomson Avenue, Cambridge, CB3 0HE, UK}

\author[0000-0003-2919-7495]{Christina C. Williams}
\affiliation{NSF National Optical-Infrared Astronomy Research Laboratory, 950 North Cherry Avenue, Tucson, AZ 85719, USA}

\author[0000-0001-9262-9997]{Christopher N. A. Willmer}
\affiliation{Steward Observatory, University of Arizona, 933 N. Cherry Avenue, Tucson, AZ 85721, USA}

\author[0000-0002-7595-121X]{Joris Witstok}
\affiliation{Cosmic Dawn Center (DAWN), Copenhagen, Denmark}
\affiliation{Niels Bohr Institute, University of Copenhagen, Jagtvej 128, DK-2200, Copenhagen, Denmark}



\begin{abstract}

We present deep JWST/NIRCam and MIRI imaging of Ion1, a previously confirmed Lyman Continuum (LyC)-emitting galaxy at $z_{\rm{spec}}=3.794$. Together with existing HST imaging, these new observations from the JADES program enable a joint analysis of Ion1's LyC, rest-frame UV, stellar, and dust emission with unprecedented detail. We report the first detection of dust emission at rest-frame $\sim3$ $\mu$m in a high-redshift LyC-emitting galaxy using MIRI/F1500W. Our analysis suggests a porous distribution of dust in Ion1, with regions exhibiting evidence of dust deficit coinciding both with LyC-emitting regions and with the peak of H$\alpha$ emission. Furthermore, multi-band NIRCam imaging reveals a strong FUV-to-optical color gradient, where LyC-emitting regions appear significantly bluer than the rest of Ion1. Spatially resolved SED fitting confirms that this color gradient is primarily driven by spatially varying dust attenuation. Together, these findings suggest that Ion1’s LyC emission originates from a compact star-forming complex near its stellar-light centroid, where stellar feedback carves out low \ion{H}{1} column density channels, facilitating LyC escape. However, only a fraction of these LyC photons -- specifically those along sightlines with minimal \ion{H}{1} obscuration -- ultimately escape and reach observers. This work underscores the critical role of dust and neutral gas geometry in shaping LyC escape in galaxies at high redshifts. Anisotropic LyC escape may be a common feature in the early Universe, which must be properly incorporated to constrain the Epoch of Reionization.

\end{abstract}


\section{Introduction} \label{sec:intro}

The sources responsible for producing the bulk of the ionizing radiation, or Lyman Continuum emission (LyC, $\lambda_{\rm{rest}}<912$ \AA), which re-ionized the Universe at redshift $z\sim5-7$ \citep[e.g.,][]{Planck2016,Bosman2022,Zhu2024} and kept it ionized ever since, remain elusive \citep{Robertson2022}. Unfortunately, directly detecting LyC emission from normal galaxies (i.e., non-quasars) is virtually impossible at the Epoch of Reionization (EoR, $z>5$) owing to the \ion{H}{1} opacity of the intergalactic medium (IGM) \citep[e.g.,][]{Madau1995,Vanzella2010,Steidel2018}.  

Significant efforts have thus been devoted to searching for and characterizing LyC-emitting galaxies at lower redshifts. With current instruments, two redshift windows have been intensively studied. The first is the local Universe of $z<0.4$, where the far-ultraviolet (FUV) sensitivity of the Cosmic Origins Spectrograph (COS; \citealt{Green2012}) aboard the Hubble Space Telescope (HST) has enabled a surge in LyC detections over the past decade \citep[e.g][]{Leitet2013,Borthakur2014,Leitherer2016,Izotov2016,Izotov2016b,Izotov2018,Izotov2018b,Wang2019,Flury2022,Roy2024}. 

The combination of deep ground-based spectroscopy and high-angular-resolution HST imaging in the rest-frame UV opens another redshift window, specifically $2<z<4$, for identifying robust LyC-emitting galaxies \citep{Vanzella2012,deBarros2016,Vanzella2016,Shapley2016,Bian2017,Steidel2018,Vanzella2018,Rivera-Thorsen2019,Fletcher2019,Ji2020,Wang2023,Gupta2024,Kerutt2024}. This range represents the highest redshift at which direct detections of LyC emission from galaxies are still possible. Consequently, galaxies confirmed to emit LyC within this redshift range are arguably the best analogs for inferring the physics associated with LyC escape during the EoR. However, it is important to note that such inferences may be affected by systematic uncertainties due to the rapid evolution of galaxy properties in the early Universe.

Among confirmed LyC-emitting galaxies at $z>3$, Ion1 is remarkable. It was originally selected as a $z\sim4$ (B-band dropout) Lyman-break galaxy in the GOODS-S field \citep{Giavalisco2004}. Ion1 was first identified as a candidate LyC emitter by \citet{Vanzella2010} using ultradeep U-band imaging from the Very Large Telescope (VLT) where its LyC was detected with a signal-to-noise ratio of S/N $\sim$ 7. Ion1 was not detected in the 7-megasecond Chandra Deep Field-South X-ray imaging \citep{Luo2017}, providing no evidence of bright active galactic nucleus (AGN) presence. Later, \citet{Ji2020} confirmed Ion1 as a robust LyC-emitting galaxy at $z_{\rm{spec}}=3.794$ using deep VLT/VIMOS spectroscopy and HST/WFC3 UVIS imaging at F410M, which directly probes Ion1’s LyC emission (i.e., LyC imaging).  

The HST LyC imaging shows that the centroid of Ion1's LyC emission is spatially offset from the non-ionizing UV and stellar emission by $\sim0.8$ kpc. The absolute escape fraction of Ion1 was measured to be $f_{\rm{esc}} = 5 - 11\%$, which is  higher than the $\sim2.0\%$ predicted by the UV slope ($\beta$)-$f_{\rm{esc}}$ relation from \citet[][their Equation 11]{Chisholm2022}\footnote{We note that new multivariate predictors of LyC escape have been proposed by \citet{Jaskot2024a, Jaskot2024b}. Using the R50–$\beta$ model (Table 24 in \citealt{Jaskot2024b}), we obtain a predicted $f_{\rm esc} = 4.5\%$ for Ion1, which is similar to the value obtained using the predictor of \citet{Chisholm2022}. For other multivariate predictors in \citet{Jaskot2024b}, they require measurements of optical spectral properties, such as $O_{32}$, which are currently unavailable for Ion1.}. Moreover, Ion1 exhibits Ly$\alpha$ absorption in the relatively low resolution ($R\sim600$) VIMOS spectrum  -- a spectral feature that is entirely different from all other known LyC-emitting galaxies which show strong Ly$\alpha$ emission. Understanding the mechanisms driving LyC photon escape in galaxies like Ion1 is therefore crucial for a comprehensive characterization of the full range of physical properties exhibited by LyC emitters.

Here, we present the new imaging observations of Ion1 taken with the James Webb Space Telescope (JWST, \citealt{Gardner2023}). Together with HST images, the deep JWST images taken from near-infrared to mid-infrared allow us to analyze, in unprecedented detail, the spatial distributions of LyC, rest-frame UV, stellar, and dust emission in Ion1 -- something that was previously impossible for such galaxies beyond $z\sim0$. Throughout this work, we adopt the AB magnitude system \citep{Oke1983} and the $\Lambda$CDM cosmology with \citet{Planck2020} parameters, i.e., $\Omega_m = 0.315$ and $\rm{h = H_0/(100\, km\,s^{-1}\,Mpc^{-1}) = 0.673}$. 

\section{Observations and Data Sets}\label{sec:data}

\begin{figure*}
    \includegraphics[width=1\textwidth]{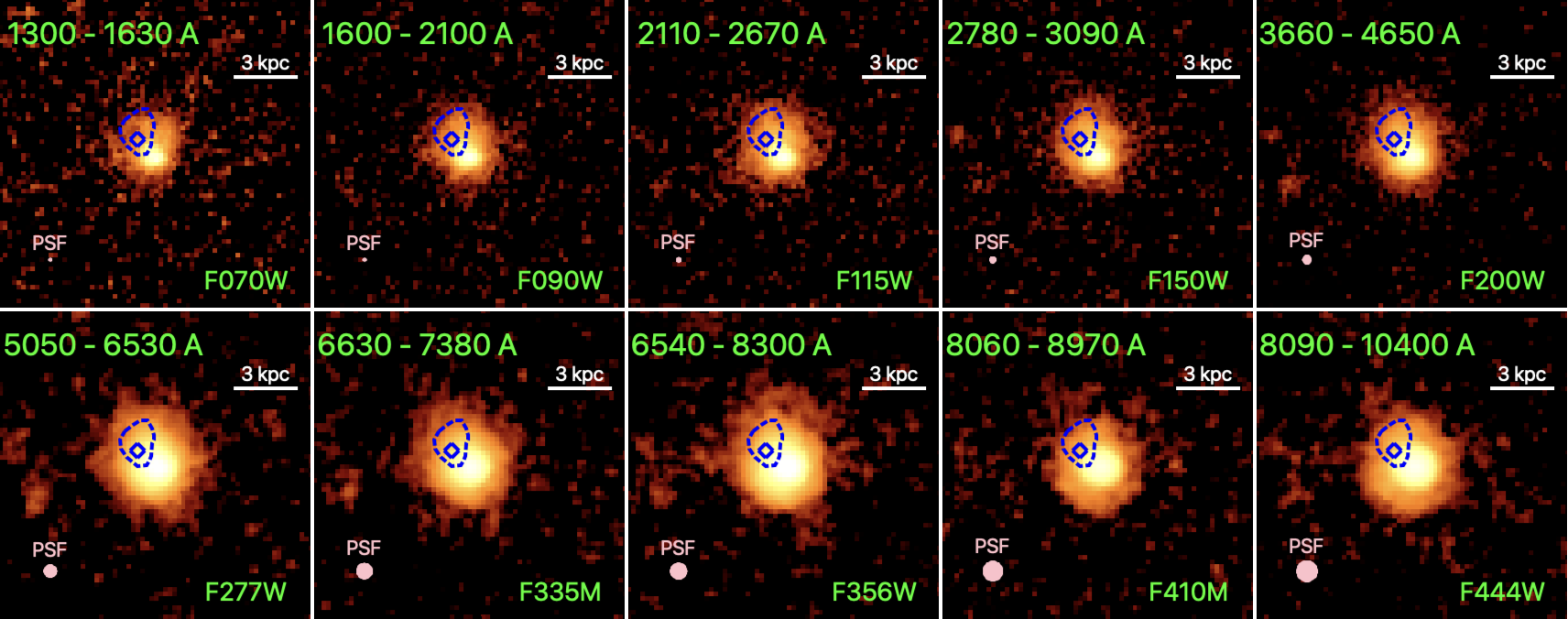}
    \caption{NIRCam 2''$\times$2'' cutouts of Ion1. The rest-frame wavelength range probed by each filter is labeled in the top-left corner. In the bottom-left corner, the angular resolution of each filter is illustrated with a filled pink circle, whose diameter is set to the FWHM of the PSF. The dashed blue contour and diamond show the 0.5$\sigma$ isophote and the peak (S/N $\sim$ 4.2) of Ion1's LyC emission (from HST WFC3/UVIS imaging obtained by \citealt{Ji2020}, see text). In these cutouts, we use a logarithmic color scale, with the minimum (black) and maximum values being the 1$\sigma$ and 200$\sigma$ local surface brightness limits. The 1$\sigma$ limits are 25.8, 26.2, 26.3, 26.4, 26.5, 27.3, 27.0, 27.4, 26.8, 27.1 mag$_{\rm{AB}}$/arcsec$^2$ from F070W through F444W.}
    \label{fig:nircam}
\end{figure*}

Here we present new observations of Ion1 obtained through the JWST Advanced Deep Extragalactic Survey (JADES, \citealt{Eisenstein2023}). Ion1 appears in JADES pointing \#1286 (PID), where deep NIRCam \citep{Rieke2023b} imaging was acquired using 10 filters: F070W, F090W, F115W, F150W, F200W, F277W, F335M, F356W, F410M and F444W. The data reduction procedures for these NIRCam observations have been described in detail in \citet{Rieke2023}. The NIRCam images of Ion1 are shown in Figure \ref{fig:nircam}.

Ion1 also appears in the JADES medium-deep MIRI \citep{Wright2023} parallel imaging observations associated with JADES pointing \#1180. These MIRI medium-deep parallels were initially planned to include three pointings overlapping with the JADES/MIRI ultra-deep F770W parallel observations in GOODS-S \citep{Alberts2024}. However, guide star failures led to a rescheduling of the MIRI imaging parallels, resulting in different position angles than originally intended. Fortunately, Ion1 appeared in one of the replanned pointings, where we obtained F1500W imaging with a 5 ks exposure, reaching a 5$\sigma$ point-source depth of 0.46 $\mu$Jy (24.7 AB mag). The data reduction for the JADES MIRI parallel imaging will be detailed in Alberts et al. (2025, in preparation) but is overall similar to the procedures outlined in \citet{Alberts2024}. All JADES NIRCam and MIRI images are aligned to the World Coordinate System (WCS) constructed based on Gaia Data Release 2 \citep{Gaia2018}.

In this work, we also utilize ancillary HST imaging to analyze the spatial distributions of LyC and non-ionizing UV radiation in Ion1 at high angular resolutions. Specifically, we include ACS imaging in the F435W, F606W, F775W, and F814W bands from GOODS-S, obtained as part of the latest Hubble Legacy Fields data release \citep{Illingworth2016,Whitaker2019}. Additionally, we incorporate the medium-deep WFC3/UVIS F410M image ($\lambda_{\rm{rest}}=$820 $-$ 890 \AA, LyC imaging) of Ion1, obtained through the HST Cycle 23 program (GO 14088, PI: Eros Vanzella), with a total on-source exposure of $\sim$ 36 kiloseconds (13 orbits). As detailed in \citet{Ji2020}, the observational strategies and data reduction procedures for the HST/F410M observations were carefully adjusted to mitigate the significant degradation in the charge transfer efficiency of the WFC3/UVIS CCDs caused by radiation exposure in the space environment.

Finally, we also utilize imaging and spectroscopy of Ion1 obtained with the VLT. These include (1) the ultradeep VIMOS U-band image, which covers rest-frame 700 $-$ 830\AA\ of Ion1 with a 1$\sigma$ limit of 29.8 AB mag, acquired by \citet{Nonino2009}, and (2) the deep, $R\sim600$ VIMOS spectrum with a total on-source exposure time of 20 hours, obtained as part of the VANDELS survey \citep{McLure2018}. Regarding the VANDELS spectrum of Ion1, we note that it was taken under seeing-limited conditions, with a seeing of approximately 1", which is comparable to the full extent of the galaxy (Figure \ref{fig:nircam}). This prevents us from placing meaningful constraints on the spatial information of the observed spectral features in Ion1.

\section{The SED properties of Ion1} \label{sec:sed}

Using photometry\footnote{All photometry is measured on PSF-homogenized images using a circular aperture with a radius of 0.5'' (i.e., ``CIRC6'' in the JADES catalog, \citealt{Rieke2023}), an aperture size chosen based on Ion1’s morphology in the NIRCam imaging shown in Figure \ref{fig:nircam}. Aperture corrections are then applied following the JADES procedure detailed in \citet{Rieke2023b}.} from HST, JWST/NIRCam and MIRI, we derive the physical properties of Ion1 through spectral energy distribution (SED) fitting with the software Prospector \citep{Johnson2021}. Given the limited understanding of the accurate IGM and circumgalactic medium (CGM) transmissions of LyC photons for individual galaxies \citep[e.g.,][]{Steidel2018}, the HST/F410M and F435W filters -- covering the LyC spectral range of Ion1 -- are excluded from the SED fitting process. The redshift of Ion1 is fixed to its spectroscopic value, i.e. $z_{\rm{spec}}=3.794$, for the SED fitting.

\subsection{Assumptions}
The basic setups, including parameter priors, in our SED modeling are essentially the same as those used in \citet{Ji2024jades_jems}. In summary, we employ the Flexible Stellar Population Synthesis code \citep{Conroy2009,Conroy2010} with the stellar isochrone libraries MIST \citep{Dotter2016,Choi2016} and the stellar spectral libraries MILES \citep{Falcon-Barroso2011}. We adopt the \citet{Kroupa2001} initial mass function, the \citet{Byler2017} nebular continuum and line emission model, and the \citet{Madau1995} intergalactic medium absorption model. The stellar and gas-phase metallicities, as well as the ionization parameter, are treated as free parameters. For dust attenuation, we assume the \citet{Charlot2000} two-component dust model. Dust attenuation for stellar populations older than 10 Myr follows the parameterization from \citet{Noll2009}, while additional dust obscuration is applied to stellar populations younger than 10 Myr and to nebular emission, following a power-law attenuation law. For dust emission, we use the models of \citet{Draine2007}, with $U_{\rm{min}}$, $q_{\rm{PAH}}$ and $\gamma$ treated as free parameters during the SED fitting (see \citealt{Draine2007} for detailed descriptions of these parameters).

We do not include an AGN component in our SED modeling, as there is no strong evidence for AGN presence in Ion1 based on deep X-ray and rest-frame UV spectroscopic observations (Section \ref{sec:intro}). While the addition of new MIRI data could further constrain the presence of an AGN in Ion1 through potential AGN dust torus emission, we note that we currently have only a single-band MIRI observation, F1500W, which covers Ion1’s spectral range only up to rest-frame 3 $\mu$m. Multiple mid-infrared filters extending to longer wavelengths are required to effectively distinguish dust emission from star formation and AGN dust torus emission \citep[e.g.,][]{Donley2012, Ji2022, Lyu2024}. As shown in Figure \ref{fig:SED}, the fact that the F1500W flux is already well reproduced by the \citet{Draine2007} model, which accounts solely for dust emission from star formation, suggests that current evidence for AGN presence in Ion1 is weak.

Regarding star-formation history (SFH), our fiducial model uses the nonparametric form with the continuity prior. The SFH is composed of 9 lookback time bins, where the first two bins are fixed to be 0 $-$ 30, and 30 $-$ 100 Myr to capture the recent star formation activity; the last bin is assumed to be 0.85t$_{\rm{H}}$ $-$ t$_{\rm{H}}$ where t$_{\rm{H}}$ is the Hubble time of $z=3.794$; the remaining 6 bins are evenly spaced in logarithmic space between 100 Myr and 0.85t$_{\rm{H}}$. We tested the impact of the assumed SFH on our conclusions by using a nonparametric SFH with different lookback time bins. Specifically, we employed finer bins ($0$–$10$, $10$–$30$, and $30$–$100$ Myr) to capture recent star formation with higher time resolution (black dashed line in the right panel of Figure \ref{fig:SED}). Additionally, we also tested our results using a parametric delayed-tau SFH (blue dashed line in the right panel of Figure \ref{fig:SED}). No substantial changes were found in our measurements.

We note that, in addition to altering the SFH assumptions in our Prospector fitting, we tested the fitting results using an alternative SED fitting framework BAGPIPES \citep{Carnall2018} with the BC03 \citep{Bruzual2003} stellar synthesis code. No substantial differences were found in our measurements. 

Finally, we tested our SED fitting results by allowing a non-zero $f_{\rm esc}$ for Ion1. To do so, in our SED fitting we add an additional FSPS parameter -- $f_{\rm obrun}$, with a uniform prior range from 0 to 1 -- which represents the fraction of young ($<$10 Myr) stars that are not attenuated by the neutral envelopes of their birth clouds and do not contribute to nebular emission. This parameter effectively accounts for escaping ionizing radiation. We obtain a best-fit $f_{\rm obrun}=0.45\pm0.15$, which is consistent with Ion1 being an
LyC leaker. Relative to our fiducial SED fitting, we find consistent stellar mass and age for Ion1 within uncertainties; the SFR increases by a factor of $\sim2$, as this SED model implies that a fraction of the ionizing photons escape, thereby requiring a higher intrinsic SFR to account for the observed nebular emission. Although similar SED-fitting analyses have been performed in recent studies \citep[e.g.,][]{Tacchella2023,Curtis-Lake2023}, we note that there is still a lack of systematic studies evaluating the consistency between SED-inferred and directly measured $f_{\rm esc}$. Moreover, we caution that converting parameters like $f_{\rm obrun}$ to $f_{\rm esc}$ is non-trivial as LyC escape is governed by small-scale physical conditions that
cannot be adequately captured in the simplified setups of SED fitting. Given that most physical properties (except SFR) of Ion1 are not sensitive to the assumed $f_{\rm esc}$ in the SED modeling, we adopt the results from the fiducial model for the analysis presented in this work.

\subsection{Inferred SED properties}

\begin{figure*}
    \centering
    \includegraphics[width=0.547\textwidth]{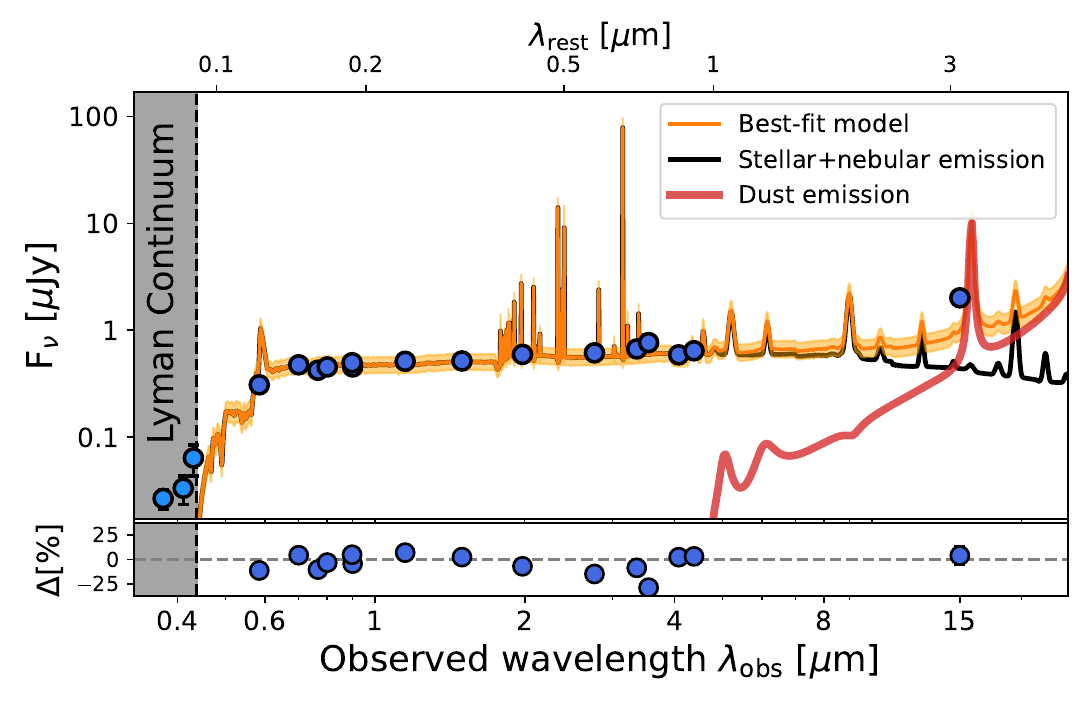}
    \includegraphics[width=0.427\textwidth]{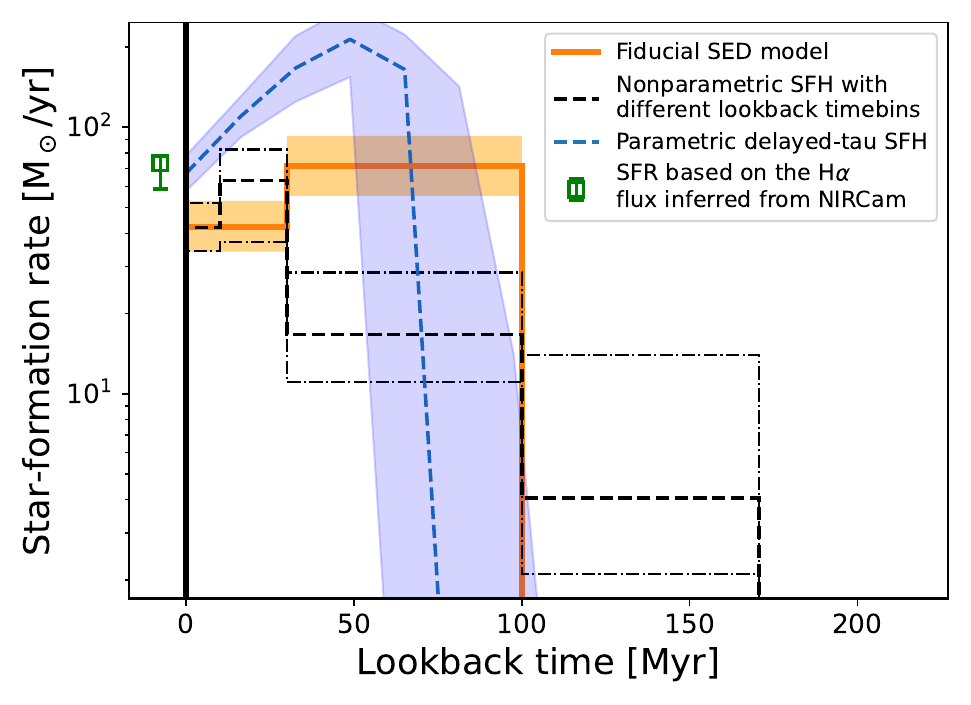}
    \caption{{\bf Left:} Fiducial SED fitting of Ion1 at $z_{\rm{spec}}=3.794$. The orange line represents the best-fit model, which is decomposed into the stellar and nebular components (black line) and the dust emission component (red line). The bottom subpanel shows the relative difference between the data and the best-fit model, calculated as (Data - Model)/Data. The three filters covering the LyC emission -- VLT/U-band, HST/F410M, and F435W (the left-most three data points in the plot) -- are excluded from the SED fitting. Hot dust emission is required to reproduce the observed F1500W flux of Ion1. {\bf Right:} Reconstructed SFH of Ion1. In addition to the SFH from our fiducial SED fitting, we also show results from two other assumed SFHs: a nonparametric continuity SFH with different lookback time bins (black dashed line), and a parametric delayed-tau ($t\cdot e^{-t/\tau}$) SFH (blue dashed line). The green square indicates the SFR estimated using the flux excess of NIRCam/F356W relative to F335M due to H$\alpha$ emission (Section \ref{sec:ha}).}
    \label{fig:SED}
\end{figure*}

Figure \ref{fig:SED} presents the SED fitting results of Ion1. Our fiducial SED modeling infers a stellar mass of Ion1 to be $\log (M_*/M_\sun) = 9.6\pm0.1$. Using different SFHs will change Ion1's stellar mass by $\pm0.3$ dex. This new stellar mass estimate is slightly higher, but in broad agreement (within $1-2\sigma$) compared to previous estimates based on HST and IRAC data \citep{Vanzella2012, Ji2020}. The reconstructed SFH suggests that the major mass assembly of Ion1 started 100 $-$ 200 Myr prior to the time of observation, which is also consistent with previous measures from \citet{Ji2020}. Table \ref{tab:info} summarizes the derived physical properties of Ion1, along with measurements from previous studies for comparison.

The MIRI data at F1500W provides brand-new constraints redward of the rest-frame 1.6 $\mu$m stellar bump of Ion1, which was entirely missed by previous studies. Ion1 is detected at F1500W with an S/N of $\sim10$ using a circular aperture of $r=0.5''$. As shown in the left panel of Figure \ref{fig:SED}, the observed F1500W flux is significantly higher than that predicted by purely stellar and nebular emission models. Dust emission is thus necessary to account for Ion1’s F1500W flux, which marks the first detection of dust emission from LyC-emitting galaxies at high redshifts. 

Since MIRI/F1500W covers the rest-frame wavelength range of $2.7–3.6$ $\mu$m for Ion1, both the thermal dust continuum and the 3.35 $\mu$m emission from the vibrational modes of Polycyclic Aromatic Hydrocarbon (PAH) molecules contribute to the observed F1500W flux. Unfortunately, with the available data, it is not possible to estimate the relative contributions of these two components. On one hand, the absence of an AGN and the high SFR in Ion1 may suggest strong 3.35 $\mu$m PAH emission relative to the underlying dust continuum \citep{Imanishi2008, Lai2020}. On the other hand, the inferred high SFR surface density (Section \ref{dis:mechanisms}) -- which can destroy PAH molecules through intense UV radiation and shocks -- and the low metallicity (Section \ref{sec:ha}) -- which can reduce the abundance of carbon-rich dust grains and thereby suppress PAH formation -- may imply a weak 3.35 $\mu$m PAH emission relative to the dust continuum \citep{Draine2007,Lai2020, Spilker2023}. Mid-infrared spectroscopy is required to determine the exact physical origin(s) of the F1500W flux from Ion1, but such data are currently unavailable. From this point forward, we refer to the F1500W flux of Ion1 simply as its dust emission at rest-frame $\sim3$ $\mu$m.

\begin{table}[]
\centering
    \caption{Physical properties of Ion1 derived in this work (fiducial SED model), along with measurements from previous studies for comparison.}
    \begin{tabular}{| c | c  | c  |}
    \toprule
           &  This work & Previous measurements $^{(\dagger)}$ \\  
    \hline 
    $\log(M_*/M_\sun)$ & 9.6 $\pm$ 0.1 $^{(*)}$ & $\in(9.0, 9.5)$ $^{a,b}$ \\
    SFR [$M_\sun$/yr] & 42 $\pm$ 5 $^{(*)}$ & $\in(18, 50)$ $^{a,b}$ \\
    E(B $-$ V) & 0.09 $\pm$ 0.05 $^{(*)}$ & $\in(0.0, 0.2)$ $^{a,b}$ \\
    $\beta_{1500}$ & -1.8 $\pm$ 0.2 $^{(**)}$ & -2.1 $\pm$ 0.2 $^{c}$ \\
    $M_{1500}$ [AB] & -22.9 $\pm$ 0.1 & -- \\
    $R_e$ [pc] $^{(***)}$& 430 $\pm$ 40 & 530 $\pm$ 60 \\
    \hline
    \end{tabular}
    
    \begin{minipage}{0.47\textwidth}
    {\small $(\dagger)$ Measurements from the following literature: (a) \citealt{Ji2020}; (b) \citealt{Vanzella2012}; and (c) \citealt{Castellano2012}. For \( M_* \), SFR, and \( E(B{-}V) \), we report the ranges derived under different SED assumptions presented in these studies. \\
    $(*)$ Adopting different SFHs changes the stellar mass by approximately \( \pm0.3 \) dex and E(B $-$ V) can increase up to $\sim0.25$ mag; the SFR increases by a factor of \(\sim 2\) when assuming a non-zero \( f_{\rm{esc}} \) in SED fitting (see Section~\ref{sec:sed}). \\
    $(**)$ UV slope is derived using HST/F775W, F814W, F850LP and NIRCam/F070W, F090W, and F115W imaging, which together cover the rest 1500--2500\AA\ for Ion1.\\
    $(***)$ UV half-light radius measured using {\sc Galfit}, assuming a single S\'{e}rsic profile. In this work, we use the NIRCam/F070W image, whereas previous measurements were based on the HST/ACS F606W image.
    }
    \end{minipage}
    \label{tab:info}
\end{table}

\section{The morphology of Ion1's Dust Emission  at rest-frame $\sim 3$ microns}\label{sec:dust_morp}

Beyond measuring the integrated flux, the exceptional angular resolution of MIRI enables a detailed analysis of the spatial distribution of Ion1's emission at F1500W. 

\begin{figure*}
    \includegraphics[width=0.97\textwidth]{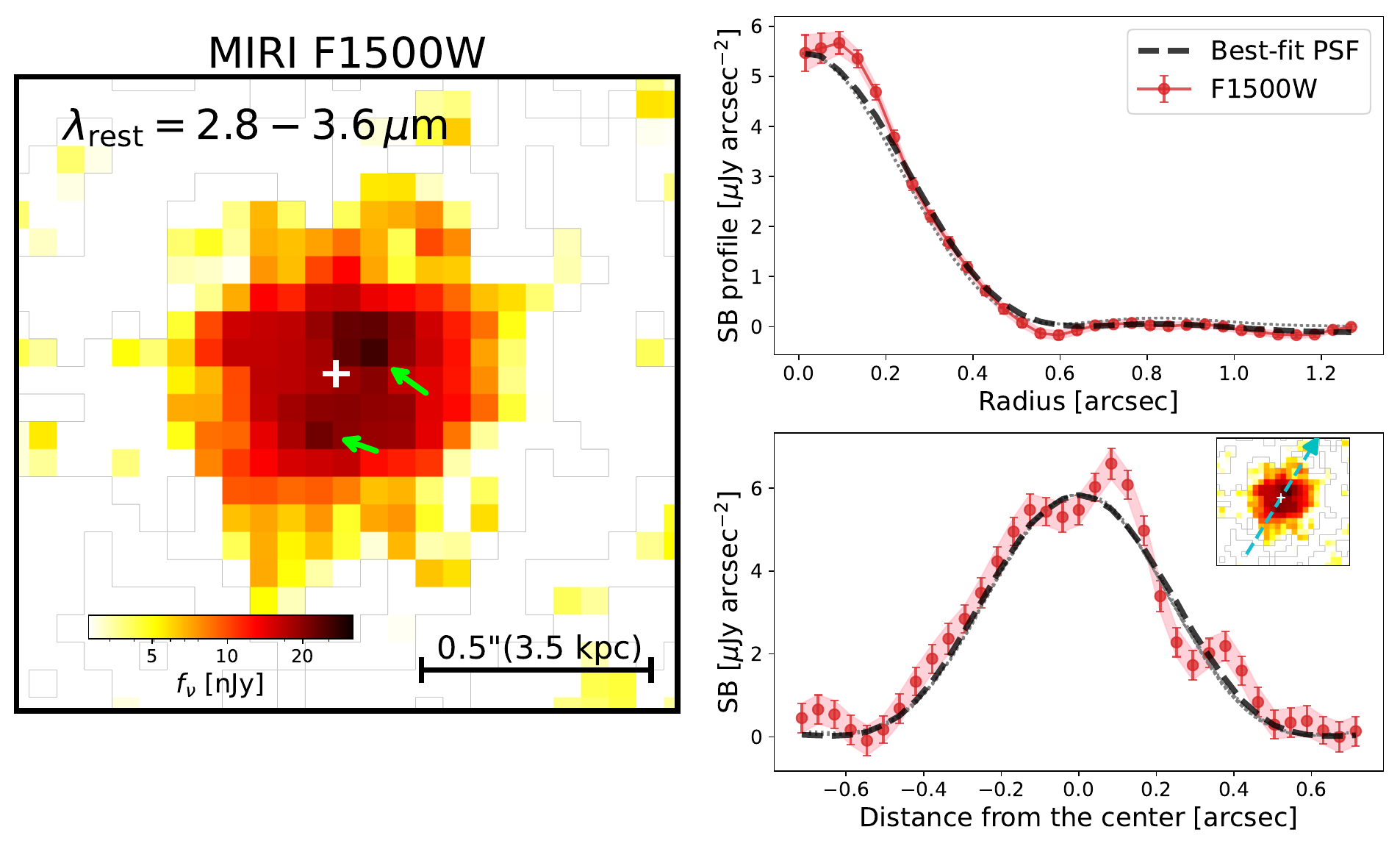}
    \caption{{\bf Left:} MIRI/F1500W imaging of Ion1. The white ``+'' marks the stellar centroid determined using LW NIRCam imaging.The light-green arrows point to the two local flux maxima referred to in the text (Section \ref{sec:dust_morp}). {\bf Right:} The top panel shows the comparison between the best-fit PSF model from {\sc Galfit} (black dashed line) and the observed, azimuthally averaged surface brightness profile (red). Similarly, in the bottom panel, we compare the best-fit PSF model with the surface brightness variation along the direction of the vector illustrated in the inset plot. In both comparisons, we also include the alternative PSF model from \citet{Libralato2024}, shown as a grey dotted line, which largely overlaps with the black dashed line, as it is very similar to the default PSF model we produced using our own F1500W mosaic (Section \ref{sec:dust_morp}). To estimate the observational uncertainty, we use the error flux extension of the MIRI data to perform Monte Carlo resampling of the imaging pixel values, repeating the surface brightness measurements 10$^4$ times. The 16th–84th percentile range is adopted as the 1$\sigma$ uncertainty. Significantly differing from an unresolved PSF, Ion1's observed hot dust (F1500W) emission shows a complex spatial distribution near the center, i.e., within $\lesssim0.2''$ of the stellar centroid.}
    \label{fig:miri_morp}
\end{figure*}

To begin, as shown in the left panel Figure \ref{fig:miri_morp}, the F1500W image shows a complex distribution of the dust emission at rest-frame $\sim 3$ $\mu$m. Apparently, two local maxima of emission -- one in the southeast direction and the other in the northwest direction -- are observed at distances of 0.1'' $-$ 0.2'' from the stellar-light centroid of Ion1, which is determined using the F444W image\footnote{The F444W image serves as a reliable proxy for the stellar light distribution of Ion1 because it probes the rest-frame $0.8-1\,\mu$m range, where no strong emission lines are present, and the stellar continuum dominates the emission.}. The complexity of the F1500W light distribution is more clearly visible in the bottom-right panel of Figure \ref{fig:miri_morp}, where we show the variation of the F1500W surface brightness along a line extending from the bottom-left to the upper-right. To best capture this variation {\it along the line}, we use a aperture with a radius of 1 pixel (0.06''), which is significantly smaller than the angular resolution of F1500W (FWHM/2 = 0.24'').

A {\sc Galfit} \citep{Peng2010} morphological analysis was also performed. We fit Ion1's F1500W image with an ePSF model constructed using bright point sources from this F1500W pointing, following the method of \citet{Anderson2000}. As shown in the top-right panel of Figure \ref{fig:miri_morp}, the PSF model reproduces the observed radial surface brightness profile reasonably well at the outskirts ($\ge$ 0.2''), but a noticeable mismatch is evident near the center, which is consistent with our surface brightness analysis presented above. We note that we also attempted to fit the F1500W image with a S\'{e}rsic profile, which yielded an unreliably small effective radius of $<0.1$ pixel. This result however is expected, as the width of Ion1’s radial surface brightness profile is comparable to the PSF width of F1500W (Figure \ref{fig:miri_morp}). 

We note that, in addition to our own ePSF model, we also tested the results of the analysis presented in this section using the F1500W PSF model from \citet{Libralato2024}, who adopted a similar method for constructing ePSFs -- namely, that of \citealt{Anderson2000} -- but combined all publicly available data taken during JWST commissioning, Cycle-1 GO, and calibration programs. As shown in Figure \ref{fig:miri_morp}, the two PSF models are highly consistent with each other, and our conclusions remain unchanged when using either PSF. Therefore, from this point onward, we report results based solely on our own F1500W ePSF.

Because the angular scale ($\sim$ 0.2'') of the complex morphological features -- specifically, the mismatch between the observed F1500W light distribution of Ion1 and the F1500W PSF at the center (Figure \ref{fig:miri_morp}) -- in Ion1's rest-frame 3 $\mu$m dust emission is comparable to the angular resolution of the F1500W imaging (FWHM/2 $= 0.24''$), it is essential to first test the null hypothesis that these features arise from noise fluctuations before proceeding further. To do this, we performed source injection simulations.

Specifically, we first created a noise-only F1500W image by masking out all sources detected at S/N $>$ 1.5 using the segmentation map generated by {\sc SExtractor} \citep{Bertin1996}. Next, we randomly injected the best-fit PSF model, derived from our {\sc Galfit} analysis, into the noise-only F1500W image. This process was repeated 10$^4$ times, and for each injection, the PSF model was resampled to include Poisson noise, calculated using the exposure time map of our MIRI/F1500W observation. Finally, we conducted the surface brightness analysis shown in Figure \ref{fig:miri_morp} for each individual injected synthetic source.

Out of 10$^4$ realizations, only 22 produced results where the surface brightness within the annular aperture of $0.1''<r<0.15''$ was (1) as bright or brighter than that observed in Ion1's F1500W image (as shown in the top-right panel of Figure \ref{fig:miri_morp}), and (2) as bright or brighter than the central surface brightness measured using a circular aperture of $r=0.1''$ (as shown in the bottom-right panel of Figure \ref{fig:miri_morp}). This corresponds to a probability of $p=0.0022$, allowing us to reject the null hypothesis with a confidence level of 99.78\%. It is worth noting that this confidence level is likely underestimated, as we reduced the information from a 2D image to a 1D profile for the sake of simplifying the quantification process of the p-value. Therefore, we conclude that the spatial distribution of Ion1's dust emission at rest-frame $\sim$ 3 $\mu$m differs significantly from that of a point source, particularly in the regions close to the center of the stellar continuum ($\le$ 0.2''). 


The observed F1500W light distribution of Ion1 could, for example, result from two or several {\it distinct} substructures, such as clumps, with overlapping isophotes. However, since no such substructures are observed at similar physical scales in the higher-resolution NIRCam images of Ion1 (Figure \ref{fig:nircam}), the F1500W morphology is mostly consistent with a porous spatial distribution of dust, namely, a smoothed distribution with certain regions (e.g., near the center of Ion1’s stellar light) exhibiting a deficit of  dust emission.

Moving forward, we will compare the morphology of the rest-frame $\sim3$ $\mu$m dust emission of Ion1 with its LyC, non-ionizing FUV, stellar and H$\alpha$ emission. Such comparisons require high astrometric precision across different images, given the small spatial scale ($\lesssim0.2''$) considered in this analysis. In Appendix \ref{app:astrometry}, we demonstrate that the relative astrometry across all images used in the analysis is robust. We will also perform pixel-by-pixel SED fitting to identify the primary factor driving the strong FUV-to-optical color gradient observed in Ion1, providing insights into the mechanisms enabling its LyC escape.

\begin{figure*}
    \includegraphics[width=0.927\textwidth]{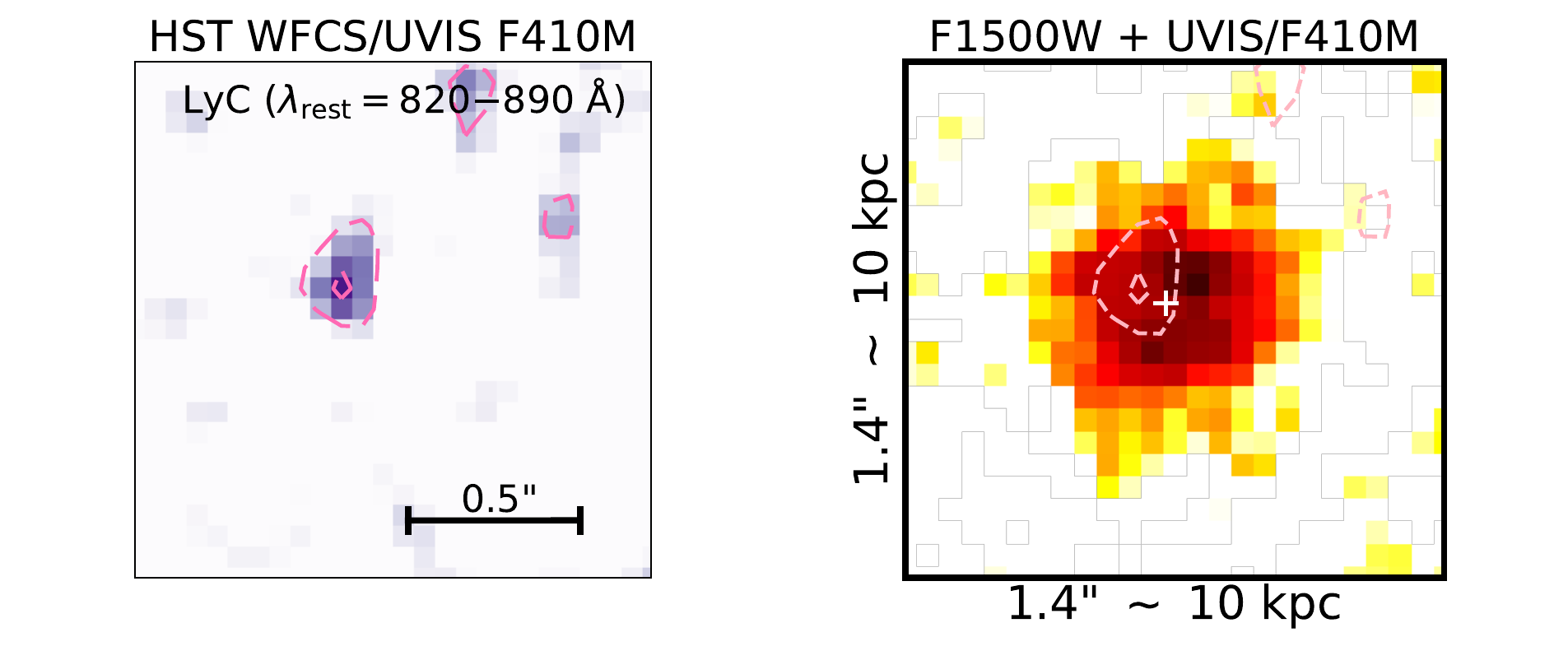}
    \includegraphics[width=0.927\textwidth]{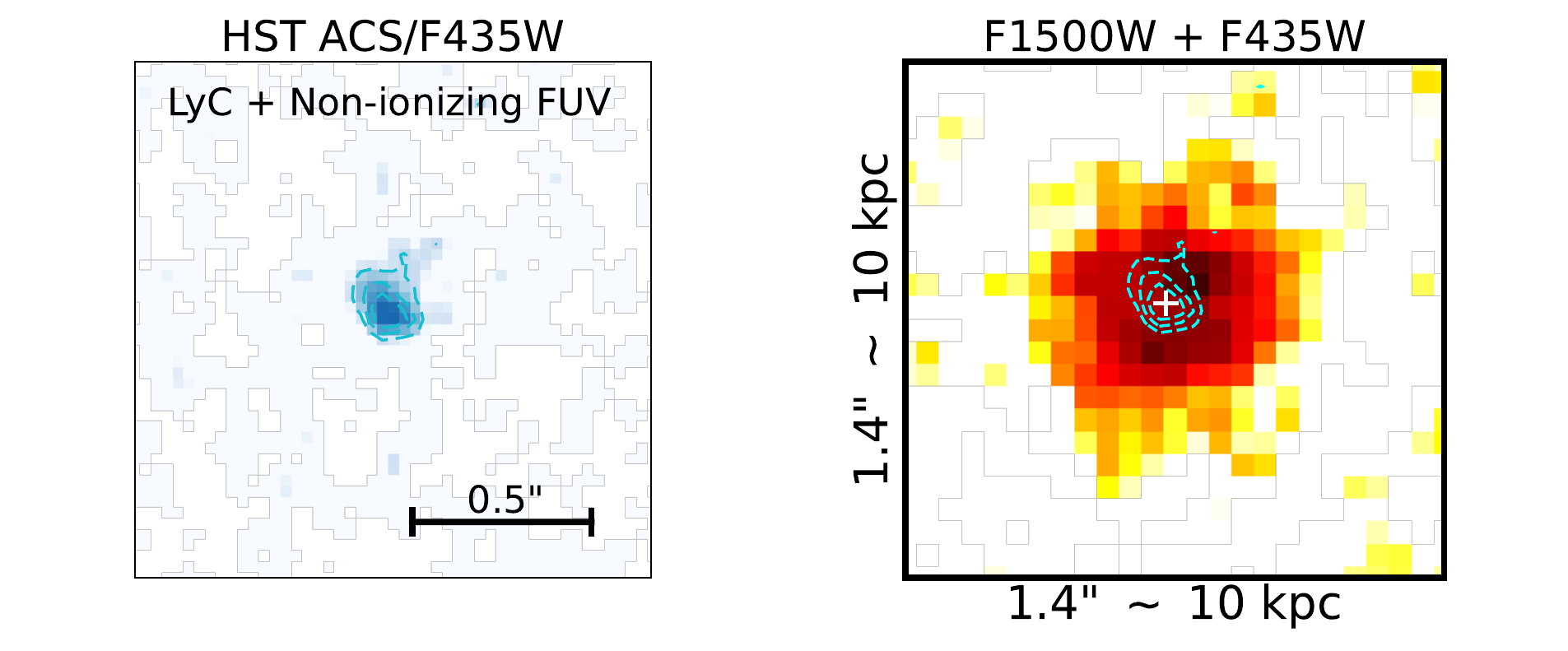}
    \includegraphics[width=0.927\textwidth]{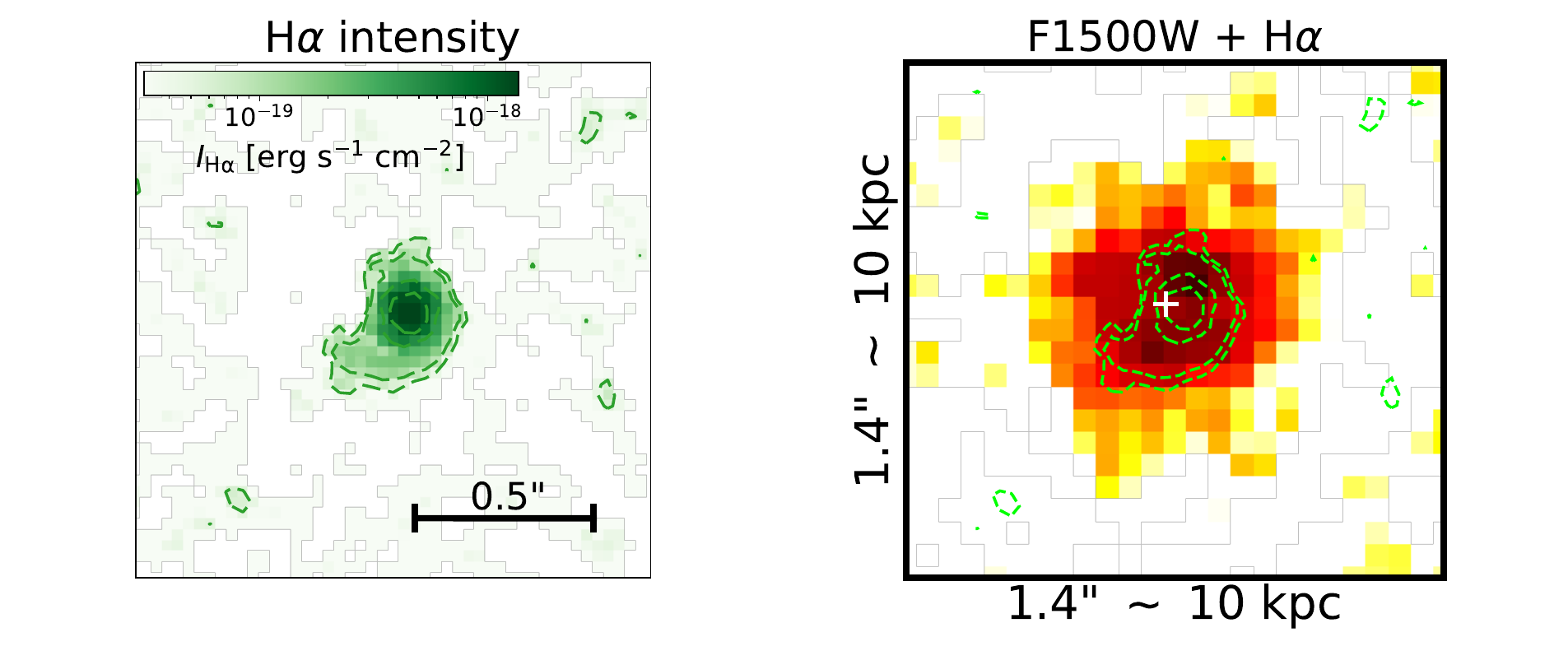}
    \caption{Comparison of the spatial distribution of the rest-frame $\sim3$ $\mu$m  dust emission (MIRI/F1500W, the same color scale as shown in Figure \ref{fig:miri_morp}) with the LyC, FUV, and H$\alpha$ emission of Ion1. From top to bottom, the left panel shows the smoothed HST/F410M (LyC,  the contours levels are the same as those shown in Figure \ref{fig:nircam}), HST/F435W (LyC+FUV, the contour levels correspond to 2$\sigma$, 3$\sigma$ and 4$\sigma$ surface brightness levels, with the 1$\sigma$ level being $\approx25.5$ mag$_{\rm{AB}}$/arcsec$^2$), and the observed H$\alpha$ map of Ion1. The H$\alpha$ map was derived using NIRCam/F356W and F335M imaging, reaching a 1$\sigma$ sensitivity of line flux per pixel at 1.9$\times10^{-20}$ erg s$^{-1}$ cm$^{-2}$ (see Section \ref{sec:ha} for details). The contour levels in the H$\alpha$ map correspond to line flux per pixel values of  5.0$\times10^{-20}$, 1.0$\times10^{-19}$, 5.0$\times10^{-19}$, and 1.0$\times10^{-18}$ erg s$^{-1}$ cm$^{-2}$. On the right, the contours from the corresponding images shown on the left are overlaid onto the F1500W image of Ion1, with the centroid of the stellar light marked by a white ``+''.}
    \label{fig:dust_vs_other}
\end{figure*}

\subsection{Comparing with LyC, FUV  and stellar emission}

\begin{figure}
    \centering
    \includegraphics[width=0.7\linewidth]{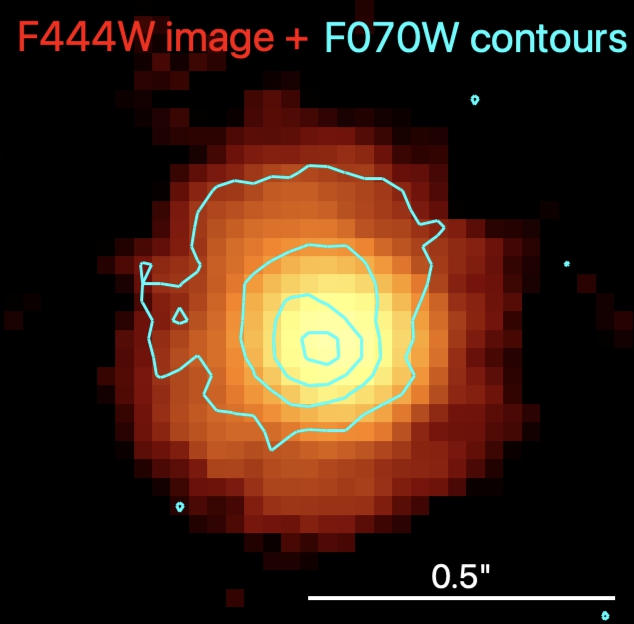}
    \caption{Comparison between the NIRCam/F070W isophotes ($\lambda_{\rm{rest}}\sim$ 1500\AA, cyan contours) and the NIRCam/F444W ($\lambda_{\rm{rest}}\sim$ 9000\AA) image of Ion1. The contours correspond to 3, 9, 27 and 81$\sigma$ levels. The rest-UV emission of Ion1 is observed preferentially from the northeast side.}
    \label{fig:uv_optical}
\end{figure}

We first compare the F1500W image of Ion1 (the dust emission at rest-frame $\sim3$ $\mu$m) with its HST/F410M and F435W images (LyC and FUV). As shown in the top and middle rows of Figure \ref{fig:dust_vs_other}, the LyC and FUV emission originate from regions {\it between} the two local maxima observed in Ion1's F1500W image, which, as interpreted above, indicates a deficit in the rest-frame $\sim3$ $\mu$m dust emission. Additionally, the LyC and FUV emission are preferentially observed from the northeast (upper-left) side of the galaxy. As shown in Figure \ref{fig:uv_optical}, the F070W image of Ion1, which does not probe the LyC or rest-FUV radiation but still probes the rest-frame UV emission at $\lambda_{\rm{rest}}\sim$ 1500\AA, also has isophotes clearly leaning toward (relative to starlight) the northeast side of the galaxy.

We note that tentative evidence of excess UV light from the northeastern regions of Ion1 was reported by \citet{Ji2020}, based on a color map produced using HST F435W and F606W images. With better-quality NIRCam imaging, we now firmly demonstrate the presence of a strong FUV-to-optical color gradient in Ion1, with the northeastern regions -- where LyC emission is observed -- appearing significantly bluer than the rest of the galaxy. This could be attributed to (1) spatially varying dust attenuation, where the northeast regions of Ion1 have less attenuation; (2) spatially varying stellar populations, such that the northeast regions produce more ionizing and FUV radiation; or (3) a combination of both factors. We will further investigate the primary driver of this strong color gradient in Ion1 using pixel-by-pixel SED fitting in Section \ref{sec:pixel_sed}.

\subsection{Comparing with H$\alpha$ emission}\label{sec:ha}

At Ion1's redshift, the H$\alpha$ emission line is observed at $\lambda_{\rm{obs}}\approx3.15$ $\mu$m, which falls within the coverage of both the NIRCam/F335M and F356W filters. However, because these two filters have significantly different transmissions at Ion1's H$\alpha$ wavelength (see Appendix \ref{app:ha}), we can utilize the two NIRCam images of Ion1 to estimate its H$\alpha$ luminosity and generate its observed (i.e., dust-uncorrected) two-dimensional H$\alpha$ intensity map. 

We detail our procedure for obtaining the H$\alpha$ map in Appendix \ref{app:ha}. In brief, the F335M image is first PSF-homogenized to match the F356W image using the JADES model PSFs \citep{Ji2024jades_jems} in the region of the sky where Ion1 appears. We estimate the continuum fluxes of Ion1 in both filters using its best-fit SED model from our fiducial Prospector fitting. We then assume the  line ratios of $I_{\rm{[NII]6584}}/I_{\rm{H}\alpha}=$ 0.32 and $I_{\rm{[NII]6584}}/I_{\rm{[NII]6548}}=$ 3,which are commonly adopted in the literature for star-forming galaxies, to estimate the contribution from the [NII] doublet to the observed flux excess in F356W relative to F335M. Finally, we estimate the H$\alpha$ flux and generate its intensity map for Ion1.

From NIRCam/F335M and F356W imaging, we estimate Ion1's total observed H$\alpha$ flux to be  $I_{\rm{H}\alpha}=(4.1\pm0.5)\times10^{-17}$ erg/s/cm$^{2}$, corresponding to an H$\alpha$ luminosity of $L_{\rm{H}\alpha}=(5.6\pm0.7)\times10^{42}$ erg/s. Using the scaling relationship between H$\alpha$ and SFR from \citet{Kennicutt2012}, we obtain the observed (uncorrected for dust) SFR$_{\rm{H\alpha}}^{\rm{obs}}$ $=30.2\pm3.7$ M$_\sun$/yr. Using the best-fit parameters of the dust attenuation towards nebular emission in our fiducial Prospector fitting, we obtain a dust-corrected SFR$_{\rm{H\alpha}}$ $=73.2_{-14.4}^{+4.5}$ M$_\sun$/yr, which is in broad agreement with our SED fitting (the right panel of Figure \ref{fig:SED}). 

We note that the estimated H$\alpha$ flux above is slightly lower than that predicted by our best-fit SED model (Section \ref{sec:sed}), which yields an H$\alpha$ flux of $(5.5\pm0.4)\times10^{-17}$ erg/s/cm$^{2}$. This discrepancy, at approximately the 2-$\sigma$ level, arises from the assumed line ratio between H$\alpha$ and the [\ion{N}{2}] doublet emission. Differing from the commonly assumed $I_{\rm{[NII]6584}}/I_{\rm{H}\alpha} = 0.32$, our SED fitting suggests a much weaker [\ion{N}{2}] emission in Ion1, resulting from its SED-inferred, substantially subsolar gas-phase metallicity of $\log(Z_{\rm{gas}}/Z_\sun) = -1.7 \pm 0.3$. In fact, if we calculate the total flux of H$\alpha$ and [\ion{N}{2}] doublet from the best-fit model and then apply the assumed $I_{\rm{[NII]6584}}/I_{\rm{H}\alpha} = 0.32$, the resulting H$\alpha$ flux would be $3.7\times10^{-17}$ erg/s/cm$^{2}$ -- fully consistent with the flux derived from NIRCam imaging. While gas-phase metallicity derived from SED fitting with photometry alone can be highly uncertain, it is noteworthy that evidence for Ion1’s low metallicity was previously reported by \citet{Ji2020}, based on a tentative constraint of the line ratio between \ion{C}{3}] $\lambda$1909 and the \ion{O}{3}] $\lambda\lambda$1661,1666 doublet emission. Accurately determining Ion1’s spatially resolved metallicity and [\ion{N}{2}] emission is beyond the scope of this work, as it requires rest-frame optical IFU spectroscopy, which is currently unavailable. However, we note that, unless the [\ion{N}{2}] doublet emission in Ion1 exhibits a dramatically different spatial distribution relative to H$\alpha$, the morphology of H$\alpha$ emission derived from NIRCam F356W and F335M imaging should closely reflect the intrinsic distribution of the observed H$\alpha$ emission.

The bottom row of Figure \ref{fig:dust_vs_other} compares the observed H$\alpha$ intensity map of Ion1 with its dust emission at rest-frame $\sim3$ $\mu$m. The centroid of the H$\alpha$ emission is co-spatial with regions exhibiting evidence of a deficit in the dust emission. However, unlike the LyC and FUV emission, whose isophotes clearly lean toward the northeastern regions of Ion1, the bulk of the H$\alpha$ emission is generally co-spatial with the stellar continuum, and the H$\alpha$ centroid is slightly offset southwest by $\sim$ 0.06''. Compared to LyC and UV emission, H$\alpha$ is less affected by dust attenuation. Therefore, the observed offset between Ion1’s H$\alpha$, LyC, and UV emission can be explained by increased dust attenuation toward the lower regions of Ion1, which preferentially absorbs UV and LyC photons more than H$\alpha$. As we discuss in Section \ref{sec:pixel_sed}, this interpretation is fully consistent with our spatially resolved SED fitting.

\subsection{Pixel-by-pixel SED analysis} \label{sec:pixel_sed}

To investigate the primary factor(s) driving the apparent differences in Ion1's light distribution across wavelengths, we perform the pixel-by-pixel SED fitting with Prospector, using PSF-homogenized NIRCam images\footnote{For this analysis, we use only the 10-filter NIRCam images from JADES, excluding HST images due to their relatively low S/N per pixel. Additionally, we do not include the MIRI/F1500W image in our current setup, as doing so would require degrading the NIRCam images to MIRI/F1500W's angular resolution, which is $\gtrsim3$ times worse than F444W. While this issue could be mitigated by forward modeling the PSF effect in pixel-by-pixel SED fitting, such an approach is beyond the scope of this work.
Nevertheless, we verified -- using the integrated photometry (Section \ref{sec:sed}) -- that including the MIRI photometry does not introduce any substantial changes to Ion1’s derived stellar population properties, including $A_V$, $\dot{n}_{\rm{ion}}$, and $\xi_{\rm{ion}}$.}, for a total of 371 pixels with S/N $>5$ at F444W. Because running Prospector is computationally expensive, for this analysis we use the {\sc Parrot} artificial neural network emulator to accelerate the fitting process \citep{Mathews2023}. We verify that the coadded stellar mass ($10^{9.7\pm0.1}M_\sun$), i.e., the sum of the stellar masses of all individual pixels, is fully consistent with the  stellar mass measurement using integrated photometry (Section \ref{sec:sed}). Additionally, the coadded SFR is $36.7\pm3.0$ M$_\sun$/yr, which is also broadly consistent with the value inferred from SED fitting using integrated photometry (shown in the right panel of Figure \ref{fig:SED}). These results indicate strong agreement between the emulator and the default Prospector framework.  A more detailed comparison for a larger sample of JADES galaxies will be presented in Zhu et al. (2025, in preparation).

\begin{figure*}
    \centering
    \includegraphics[width=0.97\linewidth]{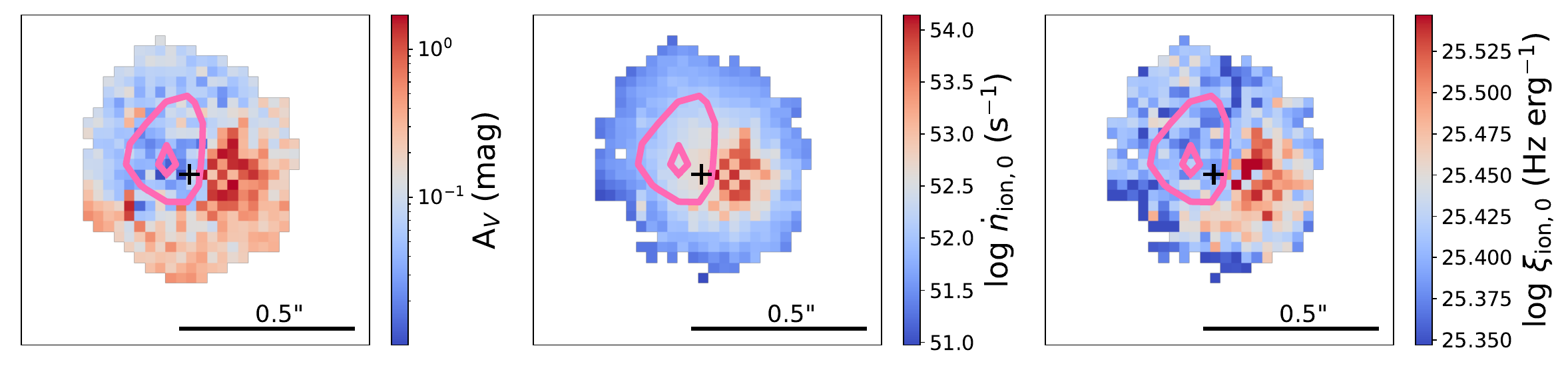}
    \caption{Pixel-by-pixel SED fitting of Ion1. From left to right, we show the maps of $A_V$, $\dot{n}_{\rm{ion,0}}$ and $\xi_{\rm{ion,0}}$, respectively. The strong FUV-to-optical color gradient of Ion1 is primarily driven by significant spatial variation in dust attenuation. Specifically, the northeastern regions of Ion1 exhibit much lower dust attenuation than the rest of the galaxy. The LyC contours, identical to those shown in the top row of Figure \ref{fig:dust_vs_other}, are overlaid in dark pink. The black ``+'' marks the center of stellar light.}
    \label{fig:pixel_sed}
\end{figure*}

We first examine the spatial variation of dust attenuation in Ion1. As shown in the left panel of Figure \ref{fig:pixel_sed}, the amount of dust attenuation, quantified by $A_V$, is significantly lower in the northeastern regions of Ion1 -- where LyC and FUV radiation are preferentially observed (Figure \ref{fig:dust_vs_other}). In principle, one could perform a detailed comparison between the spatial distributions of dust attenuation and dust emission. However, for Ion1, we only have the F1500W image, which probes the dust emission at rest-frame $\sim3$ $\mu$m. Using such dust emission to trace the total dust content remains largely uncertain and can be rather complex \citep{Spilker2023,Ji2024b}. Therefore, here we do not conduct a quantitative comparison between the 
$A_V$ map and the F1500W image of Ion1. Nonetheless, the complex spatial distribution of the dust -- suggested by both dust attenuation ($A_V$ map) and dust emission (F1500W image) -- along with the fact that the local maxima of F1500W flux do not coincide with the regions exhibiting lower $A_V$, indicate good qualitative consistency between the spatial distributions of dust attenuation and dust emission. 


Next, we examine the spatial variations of ionizing photon production rate ($\dot{n}_{\rm{ion}}$) and of ionizing photon production efficiency ($\xi_{\rm{ion}}$). Assuming Case B recombination ($f_{\rm{esc}}=0$), a temperature of 10$^4$ K, and an electron
density of 100 cm$^{-3}$, \citet{Osterbrock2006} provides the following relations
\begin{equation}
    \dot{n}_{\rm{ion,0}} = 7.28 \times 10^{11} \, L_{\rm{H\alpha}}
\end{equation}
and
\begin{equation}
    \xi_{\rm{ion,0}} = \dot{n}_{\rm{ion,0}}/L_{\rm{UV}}
\end{equation}
where $L_{\rm{H\alpha}}$ and $L_{\rm{UV}}$ are intrinsic (i.e., dust-corrected) H$\alpha$ and UV luminosities. The subscript ``0'' in $\dot{n}_{\rm{ion,0}}$ and $\xi_{\rm{ion,0}}$ emphasizes the assumption of zero $f_{\rm{esc}}$. For non-zero $f_{\rm{esc}}$, $\dot{n}_{\rm{ion,0}}$ and $\xi_{\rm{ion,0}}$ can be scaled up by a factor of $(1-f_{\rm{esc}})^{-1}$ to obtain $\dot{n}_{\rm{ion}}$ and $\xi_{\rm{ion}}$.

As shown in the middle and right panels of Figure \ref{fig:pixel_sed},  $\dot{n}_{\rm{ion,0}}$ and $\xi_{\rm{ion,0}}$ are enhanced southwest of Ion1's stellar centroid, similar to the observed spatial distribution of H$\alpha$ (Section \ref{sec:ha}). Additionally, the northeastern regions -- where the excess UV radiation is observed -- do not exhibit enhanced $\dot{n}_{\rm{ion,0}}$ or $\xi_{\rm{ion,0}}$. 

Given that LyC escape is a local phenomenon and can vary significantly within a galaxy \citep{Vanzella2012}, converting an $\dot{n}_{\rm{ion,0}}$ ($\xi_{\rm{ion,0}}$) map to an $\dot{n}_{\rm{ion}}$ ($\xi_{\rm{ion}}$) map requires a pixel-by-pixel measurement of $f_{\rm{esc}}$. However, such a measurement is not feasible for Ion1 with the current HST LyC imaging due to its limited S/N. Nonetheless, we can reasonably assume a constant $f_{\rm{esc}} = 0.1$, as reported in \citet{Ji2020}\footnote{The $f_{\rm{esc}}$ value from \citet{Ji2020} was measured only for the LyC-emitting regions in Ion1, rather than the entire galaxy.}, for the northeastern regions in Ion1 where LyC is detected (i.e., areas enclosed by the outermost 1$\sigma$ contour shown in Figure \ref{fig:pixel_sed}). For the remaining regions of Ion1, we assume $f_{\rm{esc}} = 0$, meaning $\dot{n}_{\rm{ion,0}}$ ($\xi_{\rm{ion,0}}$) $\approx$ $\dot{n}_{\rm{ion}}$ ($\xi_{\rm{ion}}$).

For the regions where LyC is observed, we calculate the median and standard deviation of $\dot{n}_{\rm{ion,0}}$ as 10$^{52.2\pm0.4}$ s$^{-1}$. Applying the correction for $f_{\rm{esc}} = 0.1$, we obtain  $\dot{n}_{\rm{ion}}= \dot{n}_{\rm{ion,0}}\times(1-0.1)^{-1} = 10^{52.3\pm0.4}$ s$^{-1}$. This value remains significantly lower than $\dot{n}_{\rm{ion}}$ found in the southwestern regions of Ion1, where H$\alpha$ is observed ($10^{53.5}$ s$^{-1}$). Similarly, for the LyC-emitting regions, the median and standard deviation of $\xi_{\rm{ion}}$ are calculated as $10^{25.45\pm0.04}$ Hz/erg, which is again lower than the corresponding value in the southwestern regions of Ion1 ($10^{25.52}$ Hz/erg).

To conclude, the pixel-by-pixel SED analysis robustly demonstrates that the strong FUV-to-optical color gradient in Ion1 is primarily driven by spatial variations in dust attenuation. The LyC photons in Ion1 likely originate from a star-forming complex near the stellar centroid, slightly offset to the southwest. These ionizing photons escape preferentially toward the northeast, where the low dust attenuation ($A_V \lesssim 0.1$ mag) creates favorable conditions for LyC escape. Specifically, the reduced dust column density in this direction implies a correspondingly lower neutral gas column density, both of which contribute to decreased absorption of ionizing photons by the ISM -- effectively forming a channel that allows for enhanced LyC emissivity.

\section{Discussion} \label{sec:diss}

\subsection{The physical mechanism of LyC escape in Ion1}\label{dis:mechanisms}
\begin{figure*}
    \includegraphics[width=0.997\textwidth]{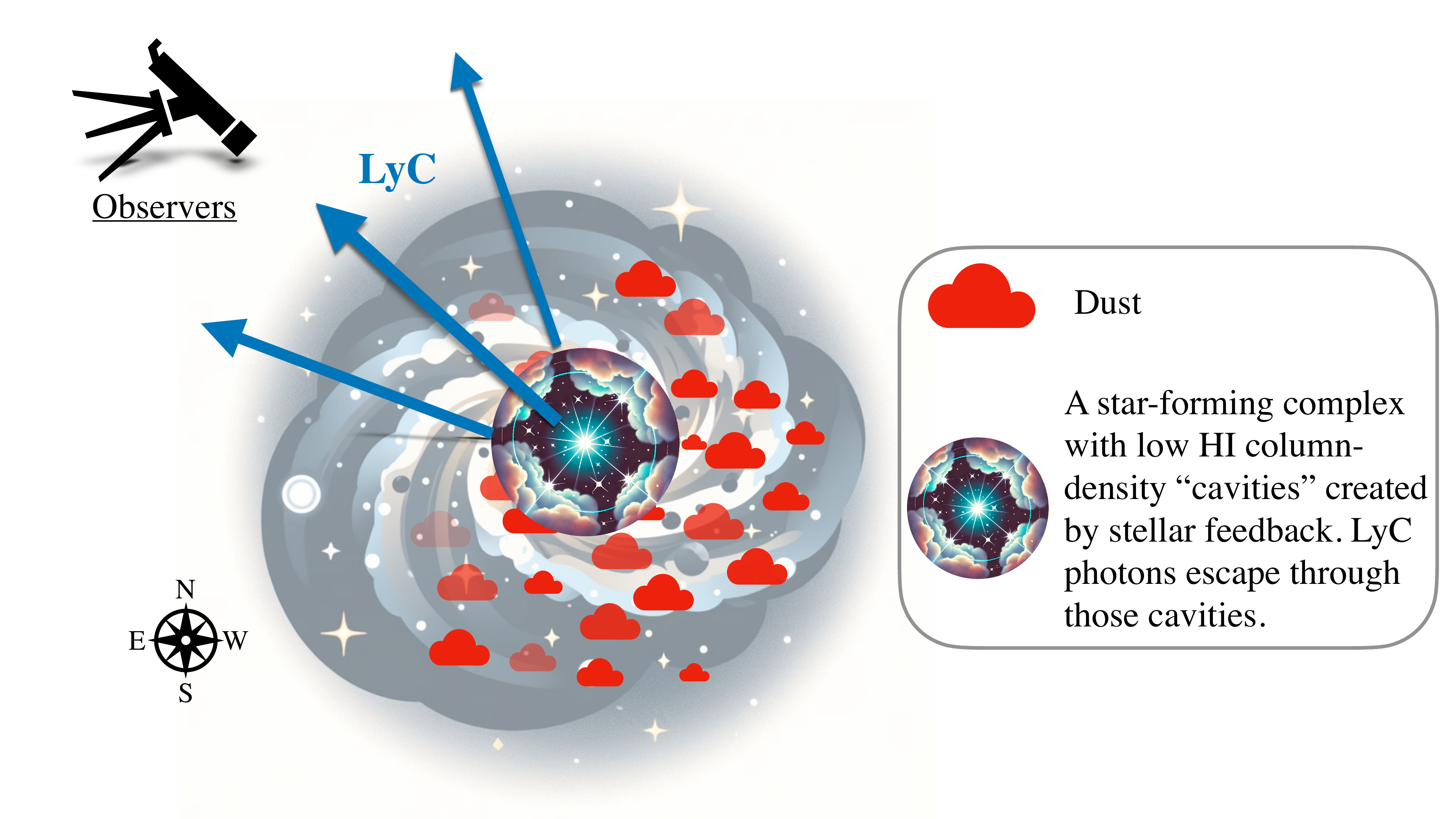}
    \caption{Illustration of LyC escape in Ion1. A detailed description is provided in Section \ref{sec:diss}. This illustration was partially created with the assistance of ChatGPT-4o \citep{OpenAI2024ChatGPT4o}.}
    \label{fig:illus}
\end{figure*}

Bringing together all findings above, we converge on a comprehensive picture of LyC escape in Ion1, as illustrated in Figure \ref{fig:illus}. Near the stellar centroid, there is a star-forming complex\footnote{With the current data, we cannot constrain the nature of this star-forming complex, which could be a single star-forming region or multiple regions in close proximity.} that dominates the ongoing star formation in Ion1. Given the absence of AGN activity in Ion1 (Section \ref{sec:intro}), this complex is likely the only primary source of LyC radiation. If low \ion{H}{1} column density cavities exist, a fraction of the LyC radiation can escape from the star-forming complex. Hydrodynamical simulations suggest that such cavities can be created by stellar feedback from highly concentrated starbursts \citep[e.g.,][]{Ma2016,Trebitsch2017,Kimm2019}. Indeed, an empirical model from \citet{Naidu2020} inferred a positive correlation between $f_{\rm{esc}}$ and the star formation rate surface density ($\Sigma_{\rm{SFR}}$), an observable sensitive to the strength of stellar feedback \citep[e.g.,][]{Heckman2001}. 

For Ion1, we derive a $\Sigma_{\rm{SFR}} = \rm{SFR_{H\alpha}}/(2\pi r_{e,\,F070W}^2) \approx 63\, M_\sun/yr/kpc^2$ \footnote{Previous studies used the half-light radius in the NUV to calculate  $\Sigma_{\rm{SFR}}$ \citep[e.g.][]{Naidu2020,Flury2022}. To maintain consistency, we use the F070W size (rest-frame 1500\AA) here.}. If we instead use the SFR from our fiducial SED fitting (Section \ref{sec:sed}) rather than that derived from the NIRCam-imaging-based H$\alpha$ flux, we obtain a $\Sigma_{\rm{SFR}} \approx 35\, \rm{M_\sun/yr/kpc^2}$. Regardless of the SFR estimation method, Ion1 exhibits a high $\Sigma_{\rm{SFR}}$\footnote{The large $\Sigma_{\rm{SFR}}$ is derived using an SFR estimated under the assumption of zero $f_{\rm esc}$. A non-zero $f_{\rm esc}$ would lead to an even larger $\Sigma_{\rm{SFR}}$ (Section \ref{sec:sed}).}, comparable to both local non-AGN-hosting strong LyC emitters \citep{Flury2022} and high-redshift Ly$\alpha$ emitters \citep{Witstok2025}. Empirically, these inferred physical conditions suggest that stellar feedback in Ion1 should be sufficient to generate low-\ion{H}{1} column density channels. Further supporting this scenario, the peak of Ion1's H$\alpha$ emission -- where $\Sigma_{\rm{SFR}}$, and thus stellar feedback, is arguably maximized -- is positioned directly between two local maxima of dust emission (Figure \ref{fig:dust_vs_other}). This alignment is consistent with a scenario in which intense stellar feedback carves out low-\ion{H}{1} channels, facilitating the escape of Ion1’s LyC photons.

Among the LyC photons streaming out from the star-forming complex, only a fraction ultimately escapes Ion1 and reaches observers -- specifically, those LyC photons traveling in the northeast direction, where additional obscuration -- originating from neutral gas {\it not immediately} associated with the star-forming complex, such as the ISM of the rest of the galaxy and the CGM -- is minimal along the line of sight.

We now discuss the potential sites where the LyC emission of Ion1 originates. The high $\Sigma_{\rm{SFR}}$ of Ion1 favors the formation of $>100$ M$_\sun$ stars within young, massive star clusters \citep{Mestric2023}. Indeed, in the case of the Sunburst Arc, where strong gravitational lensing with a magnification factor of $\mu>10$ enables HST imaging with an angular resolution of $<10$ pc, the LyC radiation is found to emerge from a single, gravitationally-bound, young ($\sim3$ Myr) star cluster with a radius of only $\sim 8$ pc \citep{Vanzella2022}. As demonstrated by \citet{Vanzella2022}, without strong lensing, this compact star cluster in the Sunburst Arc would appear as a star-forming clump with a sub-kpc size. This is remarkably similar to the observed UV size of Ion1, which is approximately 430 pc at rest-frame 1500\AA\ (NIRCam/F070W). Therefore, Ion1 may represent a similar situation to the Sunburst Arc, where its LyC radiation is primarily produced by hot, very massive stars likely confined within young, massive star clusters with sizes on the order of $\sim10$ pc.

Finally, we observe that the geometry of the LyC-emitting regions and the rest-frame $\sim3$ $\mu$m dust emission in Ion1 -- as traced by the MIRI F1500W image -- is reminiscent of a galactic fountain, similar to the archetypal case observed in the local galaxy M82\footnote{\url{https://en.wikipedia.org/wiki/Messier_82}}. This is evinced by the displacement and direction of the LyC emission relative to the centroid of the non-ionizing UV emission, which appears approximately perpendicular to the line connecting the two local maxima of the rest-frame $\sim3$ $\mu$m emission (Figure \ref{fig:dust_vs_other}). In this picture, these local maxima may trace the location of a gaseous disk in Ion1 where large amounts of dust would reside -- containing large amounts of PAH molecules and/or of warm/moderately hot dust \citep{Bolatto2024} -- while the LyC-emitting regions lie along an axis emerging from the galactic center, approximately perpendicular to this putative disk and extending above its plane. Although this interpretation remains highly speculative at present, the resemblance to a galactic fountain is noteworthy, particularly in light of ongoing efforts to understand the nature -- e.g., geometry -- of the channels that enable the escape of ionizing radiation from star-forming galaxies at high redshift.

\subsection{Consistency with the spectral features in Ion1's VIMOS spectrum} \label{sec:lya}

The picture of LyC escape in Ion1, revealed by our new NIRCam and MIRI imaging observations, now provides an explanation for the spectroscopic features observed in its VLT/VIMOS spectrum. First, as detailed in \citet{Ji2020}, residual flux is detected at the central wavelengths of several low-ionization ISM absorption lines in Ion1 (e.g., \ion{C}{2}$\lambda$1334, \ion{O}{1}$\lambda$1302, and \ion{Si}{2}$\lambda$1304), suggesting a partial covering fraction of neutral gas \citep[e.g.,][]{Mauerhofer2021}. This is consistent with the porous neutral ISM distribution indicated by both Ion1’s MIRI imaging (Figure \ref{fig:dust_vs_other}) and its pixel-by-pixel SED fitting enabled by NIRCam imaging (Figure \ref{fig:pixel_sed}).

Moreover, if we assume that the Ly$\alpha$ emission in Ion1 originates from the same region as H$\alpha$, our SED fitting suggests an attenuation of $A_V=0.1-1$ mag for Ly$\alpha$ (Figure \ref{fig:pixel_sed}). Using the $A_V$-to-$N_{\rm{HI}}$ conversion from \citet{Guver2009}, this corresponds to $N_{\rm{HI}} \approx 0.2-2 \times10^{21\,}\rm{cm}^{-2}$. In principle, $N_{\rm{HI}}$ could also be estimated directly from the observed Ly$\alpha$ profile in the VIMOS spectrum. Unfortunately, the relatively low S/N of the spectrum prevents us from placing tight constraints on $N_{\rm{HI}}$. Nonetheless, an order-of-magnitude estimate can still be obtained by comparing the observed Ly$\alpha$ profile of Ion1 with model predictions. 

\begin{figure}
    \centering
    \includegraphics[width=0.47\textwidth]{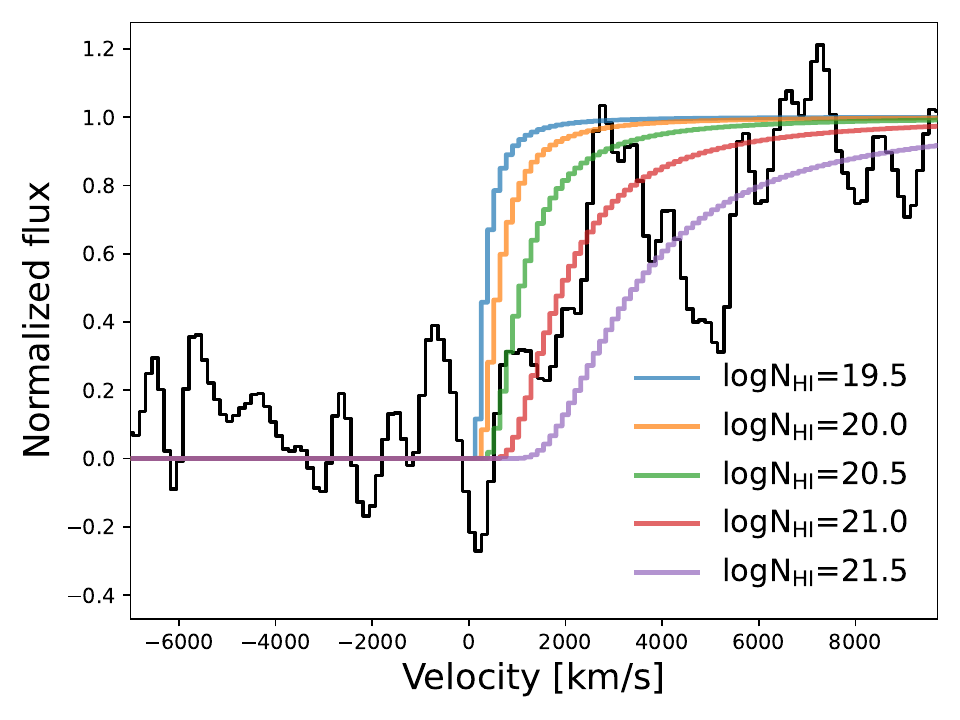}
    \caption{Comparison of model Ly$\alpha$ profiles with different $N_{\rm{HI}}$ values and the observed Ly$\alpha$ profile of Ion1 (black line), extracted from a seeing-limited, $R\sim600$ VIMOS spectrum. Given the limited S/N of the VIMOS spectrum, we do not perform detailed model fitting. Instead, our goal is to qualitatively compare the data with various models to obtain a rough estimate of $N_{\rm{HI}}$ (Section \ref{sec:lya}). Models with $N_{\rm{HI}} \in [10^{20.5}, 10^{21}] \,\rm{cm^{-2}}$ reproduce the observed profile reasonably well.}
    \label{fig:lya}
\end{figure}

Following \citet{Bolton2007}, we generate Ly$\alpha$ profiles for different $N_{\rm{HI}}$ values by modeling the optical depth of Ly$\alpha$ as an ensemble of \ion{H}{1} atoms along the line of sight -- effectively summing a series of Voigt profiles. We fix the Doppler parameter $b=10$ km/s\footnote{This is commonly used in literature. For $10^4$ K gas, one can derive  $b=\sqrt{2k_BT/m_H} \approx 13$ km/s.} and assume the Ly$\alpha$ redshift matches the systemic redshift of Ion1\footnote{Fitting the observed Ly$\alpha$ profile with the model yields a best-fit $N_{\rm{HI}}=10^{20.6\pm0.2}\,\rm{cm^{-2}}$ and a Ly$\alpha$ redshift offset of $\Delta v=-120\pm170$ km/s from the systemic redshift of Ion1.}. As shown in Figure \ref{fig:lya}, models with $N_{\rm{HI}} \in [10^{20.5},\,10^{21}]\,\rm{cm^{-2}}$ reproduce the observed profile reasonably well, in good  agreement with the SED-inferred $N_{\rm{HI}}$. Thus, the new JWST observations presented in this work also provide a natural explanation for the coexistence of LyC emission and Ly$\alpha$ absorption in Ion1. 

More stringent constraints on the relationship between the LyC and Ly$\alpha$ properties of Ion1 will require (1) an LyC map with much higher S/N and (2) spatially resolved, high spectral resolution Ly$\alpha$ spectroscopy. At $z\sim0.3$, the LaCOS survey recently obtained Ly$\alpha$ imaging for a sample of 42 galaxies with existing LyC observations from HST/COS, reporting strong correlations between Ly$\alpha$ and LyC emission when measured over small physical scales \citep{LeReste2025lacos,SaldanaLopez2025lacos}. At high redshifts, to date, the only known $z>2$ LyC-emitting galaxy that permits a detailed comparison of the spatial distributions of LyC and Ly$\alpha$ escapes is the Sunburst Arc at $z\sim2.4$ where a single LyC-emitting region is lensed into 12 distinct images, effectively sampling 12 different sightlines toward that region \citep{Rivera-Thorsen2019}. Notably, sightlines exhibiting stronger LyC emission also show higher Ly$\alpha$ emission equivalent width and escape fraction \citep{Kim2023}.
Therefore, it is possible that the region from which Ion1's LyC escapes may also be associated with a weak Ly$\alpha$ emission component. However, this emission component, if it exists, could easily be smeared out by the stronger absorption feature in a seeing-limited, low-resolution spectrum, such as the current VIMOS spectrum. 

Unfortunately, for LyC-emitting galaxies like Ion1 that are not strongly lensed, obtaining LyC imaging and IFU Ly$\alpha$ spectroscopy of the required quality is extremely challenging, if not impossible, with current facilities. The tremendous improvements in both imaging and IFU spectroscopic sensitivity at UV wavelengths offered by future instruments, such as BlueMUSE\footnote{\url{https://bluemuse.univ-lyon1.fr}} and the Habitable Worlds Observatory\footnote{\url{https://science.nasa.gov/astrophysics/programs/habitable-worlds-observatory/}}, will be critical for making such detailed studies possible.

\subsection{Tentative evidence of mergers/galaxy interactions revealed by ultradeep NIRCam imaging}

Although the results remain inconclusive, an interesting aspect has been empirically investigated by recent studies: the potentially important role of galaxy interactions and mergers in LyC escape at high redshifts \citep{ZhuS2024,Mascia2025}. At $z\sim0$, \citet{LeReste2025} recently conducted \ion{H}{1} 21 cm observations -- with a moderate resolution of $\sim$38 kpc -- for a sample of 33 galaxies with Ly$\alpha$ measurements. While no significant correlations were found between the global \ion{H}{1} and Ly$\alpha$ properties, \citet{LeReste2025} found that Ly$\alpha$ emitters are more frequently associated with signs of galaxy interactions than weak or non-emitters, suggesting that galaxy interactions may facilitate Ly$\alpha$ escape in the local Universe. With JADES imaging, which achieves a 5$\sigma$ sensitivity of $\approx30$ AB mag in all wide NIRCam filters \citep{Rieke2023}, we are able to explore the environment around Ion1 in unprecedented detail. 

\begin{figure*}
    \includegraphics[width=0.97\textwidth]{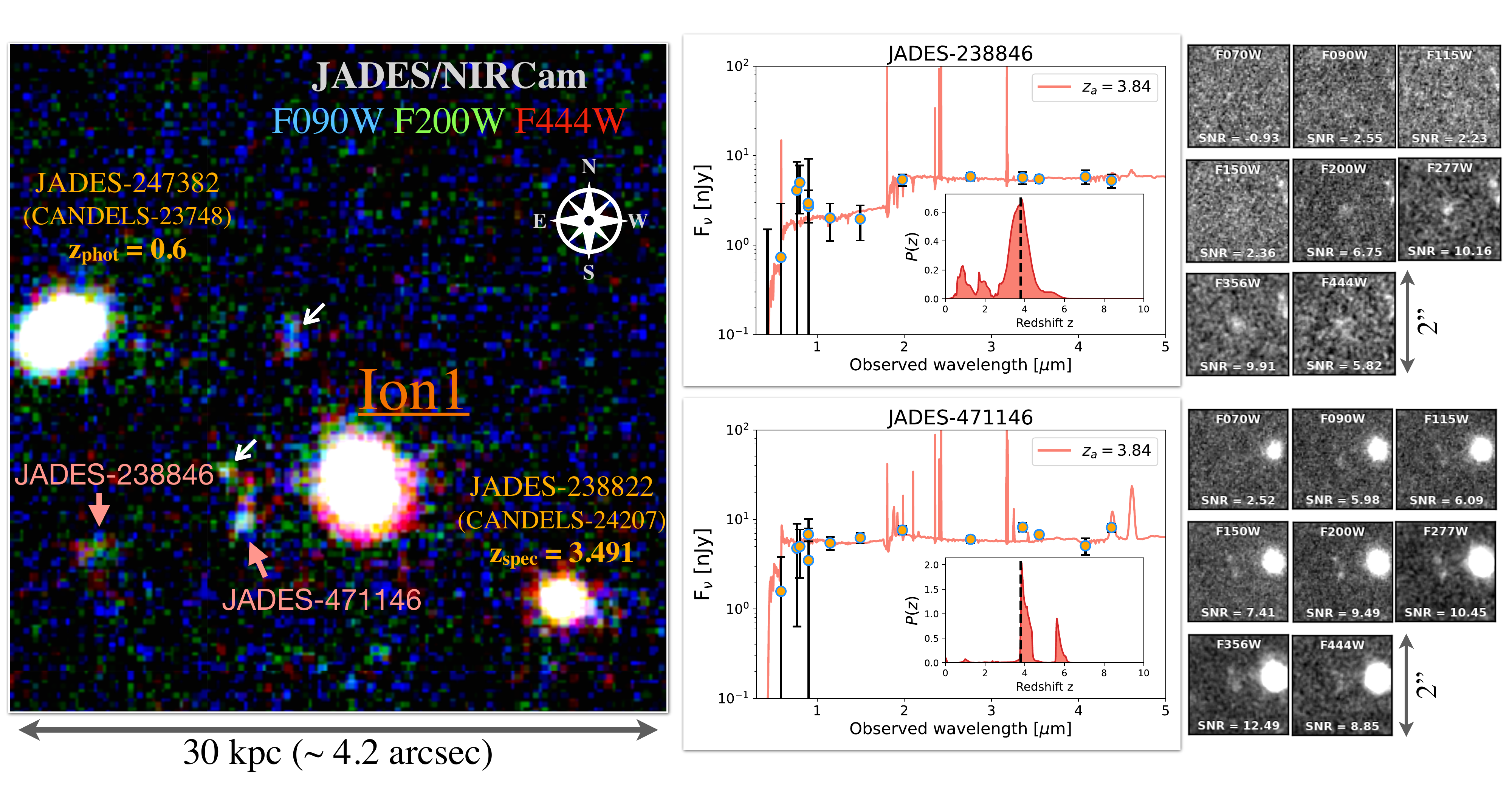}
    \caption{Deep NIRCam imaging of the regions adjacent to Ion1. Two bright foreground galaxies at $z<3.794$, previously detected by HST, are marked in orange with their redshifts and CANDELS IDs labeled \citep{Guo2013}.  White arrows indicate two galaxies newly detected by JADES/NIRCam observations, whose photometric redshifts are not consistent with Ion1. Pink arrows highlight the other two newly detected galaxies, whose photometric redshift solutions are consistent with Ion1's spectroscopic redshift. The middle panel presents the best-fit EAZY models for these two galaxies, while the inset shows the probability distribution of the photometric redshift, $P(z)\sim e^{-\chi^2(z)/2}$, with the black dashed line marking $z=3.794$. The right panel shows the NIRCam wide filter images of these two galaxies.}
    \label{fig:env}
\end{figure*}

As shown in Figure \ref{fig:env}, four new sources are detected in very close proximity to Ion1, with projected distances of $\lesssim10$ kpc at Ion1's redshift (or angular distances of $\lesssim1.4$''). These galaxies are fainter than 29 AB mag in F150W, and none were detected in the previous CANDELS HST/$H_{160}$ image \citep{Guo2013}. Following the same procedure as \citet{Hainline2024}, we perform photometric redshift analysis with EAzY \citep{Brammer2008} for these four galaxies, finding that two have photometric redshifts consistent with Ion1's spectroscopic redshift. Interestingly, both galaxies appear to the east of Ion1, i.e., in the direction where LyC is observed and dust attenuation is lower. If these two galaxies are indeed at the same redshift as Ion1, this could indicate that galaxy interactions and/or mergers facilitate LyC escape by perturbing the spatial distribution of neutral gas, creating regions depleted of neutral gas that allow LyC photons to escape more easily, a phenomenon already seen in  Haro 11, a nearby LyC emitter whose \ion{H}{1} 21cm emission shows a clear offset from its LyC emission \citep{LeReste2024}. However, we stress that, without spectroscopic confirmation, no definitive conclusions can be drawn about the impact of galaxy interactions on Ion1's LyC escape. 

\subsection{Implications for the EoR: the importance of anisotropic LyC escape}

Finally, we discuss our new constraints on LyC escape in Ion1 in the context of EoR. Among all key factors, the dominant source of uncertainty in understanding the role of galaxies in reionizing the Universe is $f_{\rm esc}$. Assuming a single value of $f_{\rm{esc}}$, a typical $f_{\rm{esc}}$ of 10\% $-$ 20\% is required for galaxies to produce enough ionizing photons to reionize the Universe by $z\sim6$ \citep[e.g.,][]{Robertson2013,Robertson2015,Finkelstein2015,Bouwens2015}. When considering the dependence of $f_{\rm{esc}}$ on galaxy mass -- primarily suggested by simulations \citep[e.g.,][]{Kimm2014, Paardekooper2015} but weakly supported by observational studies examining the correlation between stellar mass and $f_{\rm{esc}}$ for LyC-emitting galaxies \citep[e.g.,][]{Flury2022b, Fletcher2019,Saxena2022} -- and including reasonable AGN contributions, \citet{Finkelstein2019} showed that an average $f_{\rm{esc}}$ of 5\% would be sufficient for the EoR. 

Observationally, however, even an $f_{\rm{esc}}$ of 5\% is rarely detected in LyC-emitting galaxies at lower redshifts than the EoR. Moreover, stacking analyses find that non-LyC-emitting galaxies -- which constitute the vast majority of the galaxy population -- have an average $f_{\rm{esc}}$ of only $<1\%$ \citep{Fletcher2019,Ji2020}. This is puzzling, especially given that the non-LyC-emitting galaxies in these analyses share very similar physical properties with LyC-emitting ones, including redshift, mass, and $M_{UV}$.

One possible scenario 
to explain the large difference in $f_{\rm{esc}}$ between LyC-emitting and non-LyC-emitting galaxies is anisotropic LyC escape. In this picture, LyC radiation can only be detected from a galaxy if the observer’s sightline aligns with the small opening angle of its low \ion{H}{1} column density channels. Beyond the local Universe, the most direct -- and likely the clearest -- confirmation of anisotropic LyC escape is observed in the Sunburst Arc at $z\sim2.4$ \citep{Rivera-Thorsen2019,Kim2023}. Additionally, indirect evidence supporting this scenario  has also been reported, primarily from modeling of UV-to-optical spectroscopy of $z\sim0.3$ LyC emitters, which suggests partial covering of \ion{H}{1} and the presence of multiple photoionization zones in these galaxies \citep[e.g.,][]{Gazagnes2020, Ramambason2020, SaldanaLopez2022, Hu2023}.

The case of Ion1 now demonstrates that highly anisotropic LyC escape can occur at $z\sim3.8$, the highest redshift where LyC can still be directly detected from normal galaxies. Furthermore, NIRCam and MIRI imaging show that, in the case of Ion1, this anisotropy most likely arises from the complex spatial distribution of dust (and, by extension, neutral ISM) shaped by stellar feedback -- a phenomenon repeatedly found in LyC escape simulations \citep[e.g.,][]{Gnedin2008, Paardekooper2015, Trebitsch2017}. If Ion1 is representative of EoR galaxies, this suggests that anisotropy plays a crucial role in ionizing photon escape. Properly accounting for this effect is essential for understanding the physical mechanisms governing the EoR.

Unfortunately, precisely quantifying how analogous Ion1 is to galaxies in the EoR remains challenging, if not impossible. Nevertheless, Ion1 shares several key physical properties with typical EoR star-forming galaxies. First, the stellar mass of Ion1, $10^{9.6\pm0.3}M_\sun$, is very close to the latest constraints on the characteristic stellar mass ($M^\star$) of the $z\sim6$ galaxy stellar mass function \citep[e.g.,][]{Weibel2024}, where galaxies near this mass contribute the most to the total cosmic stellar mass density. For a main-sequence galaxy of this stellar mass, we estimate an SFR of $30\pm20$ M$_\sun$/yr using the star-forming main sequence from \citet{Popesso2023}, or $15^{+17}_{-8}$ M$_\sun$/yr using the star-forming main sequence derived from JADES data (Simmonds et al., 2025, in preparation). We note that using different estimators may introduce systematic uncertainties in the SFR measurement. The SFR estimators used to determine the star-forming main sequence in both studies are primarily based on SED fitting assuming zero $f_{\rm esc}$. If we adopt the SFR from our fiducial SED fitting, Ion1’s SFR is consistent with that of galaxies on the main sequence at $z\sim6$. However, when using the SFR derived from NIRCam-imaging-inferred H$\alpha$ flux or from SED fitting with a parametric delayed-tau SFH, Ion1 appears elevated above the $z\sim6$ star-forming main sequence, although not as high as typical starburst galaxies \citep[e.g.,][]{Rinaldi2022}. Additionally, Ion1's ionizing photon production efficiency $\xi_{\rm{ion}}$ (Section \ref{sec:pixel_sed})  is approximately $10^{25}$ Hz/erg, which is similar to the median value found in H$\alpha$ emitters in the EoR \citep[e.g.,][]{Rinaldi2024}.  Moreover, the median rest-optical effective radius of galaxies at $6<z<7$, measured using JADES/F444W imaging, is $\sim350$ pc (JADES collaboration, 2025, in preparation). Using the F277W image, which traces the rest-optical light of Ion1, we measure an effective radius of 370 pc, again similar to the median size of galaxies in the EoR. 

Given all these similarities, we argue that Ion1 can serve as a valuable example -- perhaps even a meaningful reference -- for understanding LyC escape in the early Universe. Notably, in $z>3$ LyC-emitting galaxies with HST LyC imaging, LyC radiation frequently appears spatially offset from non-ionizing radiation \citep{Fletcher2019, Ji2020, Kerutt2024, Gupta2024, Yuan2024}. This suggests that anisotropic LyC escape -- similar to what is observed in Ion1 -- may be a common feature of typical star-forming galaxies in the EoR. Finally, we recall that our observations are consistent with, in fact suggestive of, the possibility that the LyC emission of Ion1 originates in a galactic fountain not unlike the one observed in the local galaxy M82, as we mentioned in Section \ref{dis:mechanisms}.

\section{Summary}

We present deep JWST/NIRCam and MIRI imaging, obtained through the JADES survey, of Ion1 -- a previously confirmed LyC-emitting galaxy at $z\sim3.8$. 

With MIRI/F1500W imaging, we report the detection of dust emission at rest-frame $\sim3$ $\mu$m -- where both the thermal dust continuum and the 3.35 $\mu$m PAH emission fall within the F1500W coverage -- in Ion1, marking the first such detection in a high-redshift LyC-emitting galaxy. A detailed morphological analysis reveals a complex light distribution of Ion1 at F1500W, where two local flux maxima are observed at distances of 0.1''$-$ 0.2'' from its stellar-light centroid. Because no such substructures are observed at similar physical scales in the higher-resolution NIRCam images, we interpret the F1500W morphology of Ion1 as a smoothed distribution of light, with certain regions (e.g., near the center of Ion1’s stellar light) exhibiting a deficit of rest-frame $\sim3$ $\mu$m dust emission.

Comparing the spatial distribution of Ion1's  rest-frame $\sim3$ $\mu$m dust emission (F1500W) with its emission at shorter wavelengths, we show that regions exhibiting evidence of dust depletion coincide with those where LyC emission is observed. Moreover, the peak of Ion1's  H$\alpha$ emission, derived using the H$\alpha$ map produced from NIRCam F335M and F356W images, lies directly between the two local maxima observed in the dust emission. Additionally, deep multi-band NIRCam imaging reveals a strong FUV-to-optical color gradient in Ion1, where the northeastern regions -- where LyC is observed -- appear significantly bluer than the rest of the galaxy. Pixel-by-pixel SED fitting confirms that this color gradient is primarily driven by spatially varying dust attenuation rather than differences in stellar populations. Together, these findings provide compelling evidence that the spatial distribution of dust, and hence the neutral ISM, in Ion1 is porous, rather than a uniformly distributed dust screen.

The new NIRCam and MIRI observations reveal the underlying physical mechanism driving LyC escape in Ion1. The ionizing photons originate from a compact star-forming complex near the galaxy's stellar centroid, with a high star formation rate surface density of $\Sigma_{\rm{SFR}} = 63\, \rm{M_\sun/yr/kpc^2}$. Such a high $\Sigma_{\rm{SFR}}$ enables strong stellar feedback, which carves out low \ion{H}{1} column density channels, allowing LyC photons to escape from the complex. However, only a fraction of these LyC photons ultimately escape Ion1 and reach observers -- specifically, those traveling along the northeastern direction, where line-of-sight \ion{H}{1} obscuration is minimal. 

The picture above, revealed by our new JWST observations, also provides a natural explanation for the spectral features observed in Ion1's VIMOS spectrum, including the residual fluxes at the central wavelengths of low-ionization ISM absorption features and the presence of Ly$\alpha$ absorption.

As a closing remark, we note that this study of Ion1 highlights the crucial role of dust and neutral gas geometry in shaping LyC escape at high redshifts. Currently, similar studies do not exist for statistically significant samples of LyC-emitting galaxies. However, such constraints are urgently needed, as they provide critical insights into the mechanisms regulating LyC photon leakage and the role of star-forming galaxies in cosmic reionization. Understanding how dust and neutral gas are spatially distributed -- and how they influence LyC escape -- is essential for refining theoretical models of reionization and improving predictions of ionizing emissivity from galaxies. The study of Ion1 represents an early step in this direction, demonstrating the power of combining HST, NIRCam, and MIRI imaging to investigate LyC escape and its connection to the spatially resolved properties of galaxies. It also underscores the need for similar observations of larger samples of LyC-emitting galaxies -- particularly those at $3<z<4$, the highest redshift range where LyC emission can still be directly detected -- to systematically characterize the physical conditions that facilitate LyC escape in the early Universe. Finally, the intriguing similarity between the relative geometry of the LyC emission and the dust emission (PAH emission and/or thermal dust continuum) observed in Ion1, and that of a galactic fountain in M82 offers an exciting opportunity to further explore the nature and formation of the channels through which ionizing radiation escapes into the CGM and IGM. This could be pursued, for example, through spatially resolved spectroscopy at mid- and far-infrared wavelengths, which could help identify the putative gaseous and dusty disk -- perpendicular to which the ionizing radiation is escaping.
\\\\
\section*{acknowledgments}

ZJ, YZ, KH, JMH, BER and CAW acknowledge support from JWST/NIRCam contract to the University of Arizona NAS5-02015. SA acknowledges support from the JWST Mid-Infrared Instrument (MIRI) Science Team Lead, grant 80NSSC18K0555, from NASA Goddard Space Flight Center to the University of Arizona. WMB gratefully acknowledges support from DARK via the DARK fellowship. This work was supported by a research grant (VIL54489) from VILLUM FONDEN. AJB acknowledges funding from the ``FirstGalaxies'' Advanced Grant from the European Research Council (ERC) under the European Union's Horizon 2020 research and innovation programme (Grant agreement No. 789056). BER acknowledges support from the JWST Program 3215. CS acknowledges support from the Science and Technology Facilities Council (STFC), by the ERC through Advanced Grant 695671 ``QUENCH'', by the UKRI Frontier Research grant RISEandFALL. ST acknowledges support by the Royal Society Research Grant G125142. The research of CCW is supported by NOIRLab, which is managed by the Association of Universities for Research in Astronomy (AURA) under a cooperative agreement with the National Science Foundation. JW gratefully acknowledges support from the Cosmic Dawn Center through the DAWN Fellowship. The Cosmic Dawn Center (DAWN) is funded by the Danish National Research Foundation under grant No. 140. 

This work made use of the {\it lux} supercomputer at UC Santa Cruz which is funded by NSF MRI grant AST 1828315, as well as the High Performance Computing (HPC) resources at the University
of Arizona which is funded by the Office of Research Discovery and Innovation (ORDI), Chief Information Officer (CIO), and University Information Technology Services (UITS).

\bibliography{sample631}{}
\bibliographystyle{aasjournal}

\appendix 

\section{Astrometric precision across different images} \label{app:astrometry}

\begin{figure}
    \centering
    \includegraphics[width=1\linewidth]{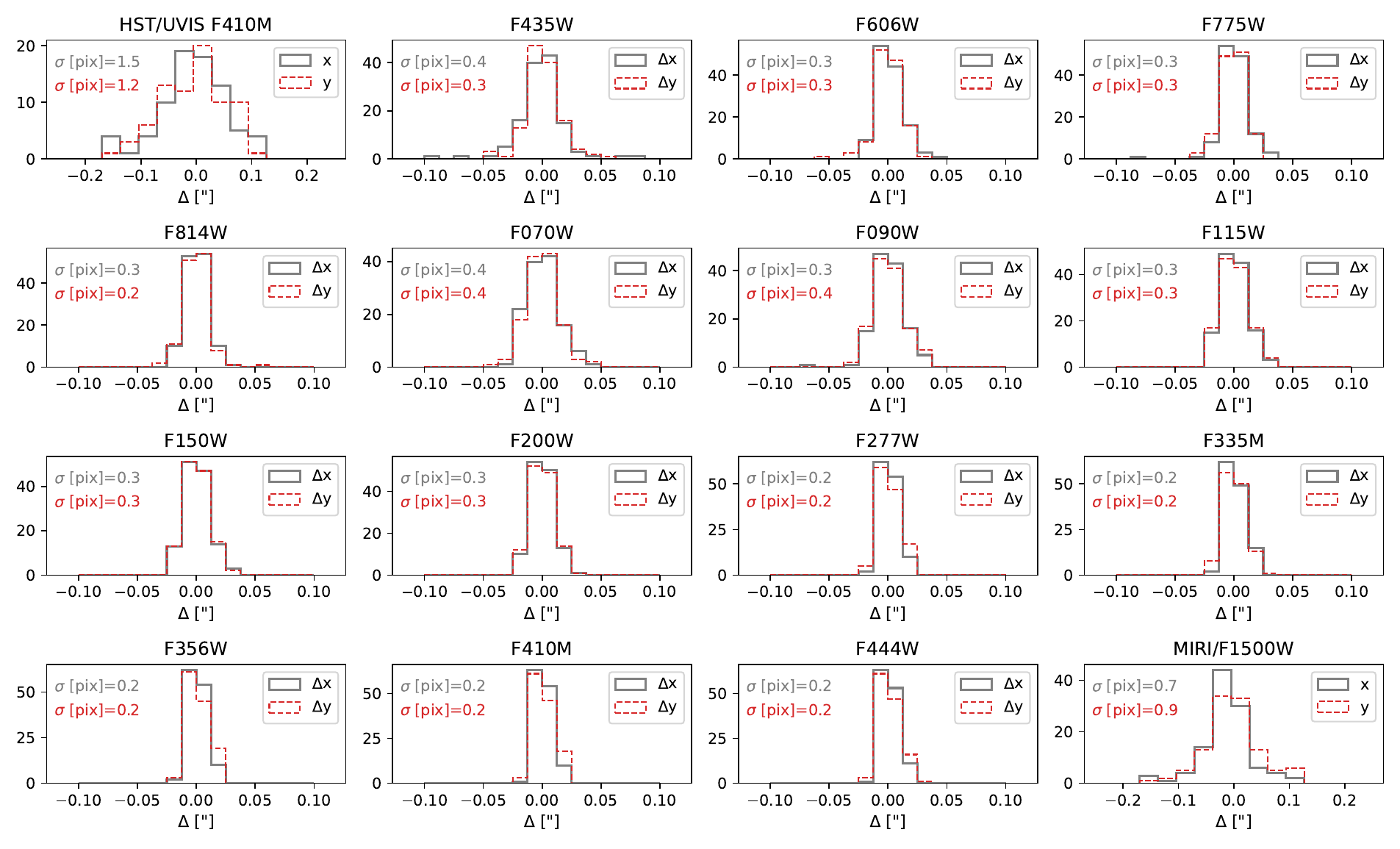}
    \caption{Astrometry comparison. Filter names are labeled at the top of individual panels. The histograms show the difference (in arcseconds) between the centroid measured in each corresponding filter and the centroid listed in the JADES photometric catalog. The black histogram represents the difference in the $x$-direction, while the red histogram represents the difference in the $y$-direction. The standard deviations (in pixels) are also labeled in each panel.}
    \label{fig:wcs}
\end{figure}

Our results rely only on relative astrometry rather than absolute astrometry. To ensure high precision in relative astrometry, we first apply astrometric corrections to all HST, JWST/NIRCam, and MIRI images within a
$d<30''$ region around Ion1 using the JADES photometric catalog. We select sources with S/N $>10$ (measured within a $r=0.2''$ circular aperture) in NIRCam/F200W, and determine their centroids in each filter using \texttt{photutils.centroids} \citep{photutils}. For each image, we then correct its WCS header by applying the median difference between these centroids and those listed in the JADES catalog\footnote{Note that source detection and segmentation were performed based on an inverse-variance-weighted stack of the NIRCam/F277W, F335M, F356W, F410M, and F444W images \citep{Rieke2023b}.}. As shown in Figure \ref{fig:wcs}, after applying these corrections, in all filters  the median astrometric differences are consistent with zero, and the standard deviations are $\lesssim1$ pixel.

\section{Deriving Ion1's H$\alpha$ intensity map using the F335M and F356W images} \label{app:ha}

\begin{figure*}
    \centering
    \includegraphics[width=0.5\linewidth]{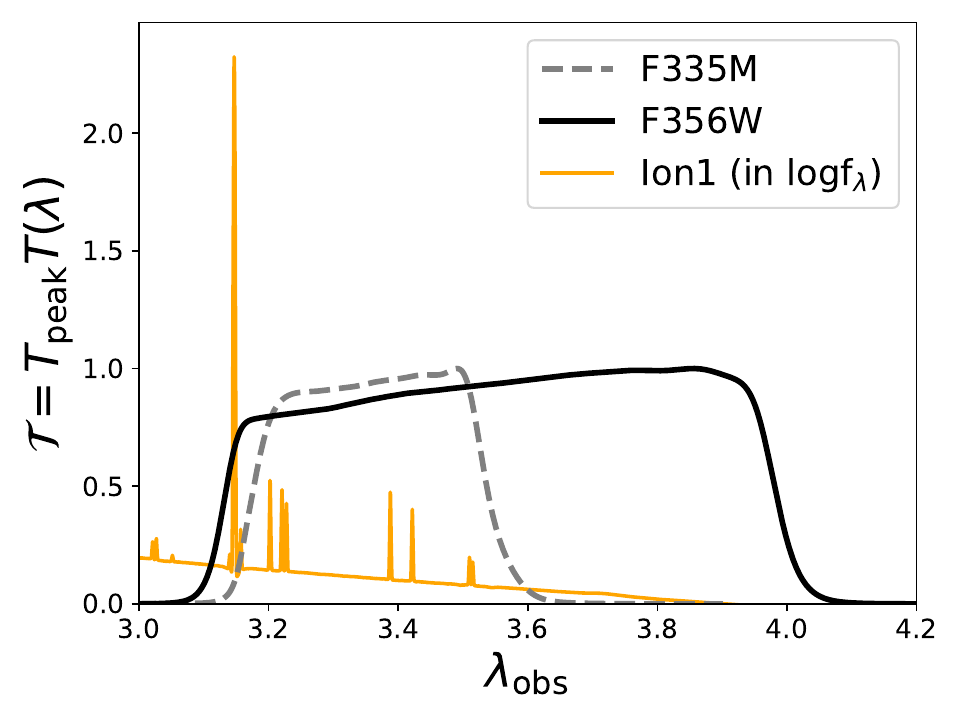}
    \caption{The H$\alpha$ of Ion1 is covered by both the F335M and F356W images, but it appears at the parts of the two filters with largely different transmissions, allowing us to derive the 2D H$\alpha$ intensity map using the two NIRCam images (Appendix \ref{app:ha}).}
    \label{fig:ha_filters}
\end{figure*}

The wavelength range where the H$\alpha$ of Ion1 appears is very close to F335M's transmission cutoff at the blue end, with a transmission (relative to the maximum transmission of the filter) of only $\approx$ 8\%, while it corresponds to the wavelength range with a $\approx$ 65\% transmission of F356W (see Figure \ref{fig:ha_filters}). 

Given the relatively large widths of the F335M and F356W filters, the line widths of H$\alpha$ and [NII] doublet are assumed to be infinitely small. The spectrum of Ion1 thus can be written as
\begin{equation}
    f_{\lambda} \approx f_{\rm{c}} + \sum_{i=1}^{3}I_{i} \delta(\lambda-\lambda_{i})
\end{equation} 
where i$=$1, 2 and 3 corresponds to H$\alpha$, [NII]6548 and [NII]6584, respectively. 
Following \citet{Pascual2007}, for a given filter $\mathcal{F}$ with a relative transmission curve of $\mathcal{T}(\lambda)$  which is normalized to the peak transmission such that the absolute T$=$T$_{\rm{peak}}$$\mathcal{T}$, the observed flux is  
\begin{equation}
    \langle f_{\lambda}^{\mathcal{F}}\rangle = \langle f_{\rm{c}}^{\mathcal{F}}\rangle + I_{\rm{H}\alpha}\cdot\frac{1}{\Delta^{''}(\lambda_{\rm{obs}}^{\rm{H}\alpha})}
    \label{equ:mean_f}
\end{equation}
where 
\begin{equation}
    \begin{split}
    \Delta^{''}(\lambda_{\rm{obs}}^{\rm{H}\alpha}) & = \left[ \frac{1}{\Delta^{'}(\lambda_{\rm{obs}}^{\rm{H}\alpha})}+\frac{r_2}{\Delta^{'}(\lambda_{\rm{obs}}^{\rm{[NII]6548}})}+\frac{r_3}{\Delta^{'}(\lambda_{\rm{obs}}^{\rm{[NII]6584}})}\right]^{-1}\\
     r_2 & = \frac{I_{\rm{[NII]6548}}}{I_{\rm{H}\alpha}}\\
     r_3 & = \frac{I_{\rm{[NII]6584}}}{I_{\rm{H}\alpha}}
    \end{split}
    \label{equ:mean_eff_wid}
\end{equation}
In the equation above, $\Delta^{'}$ is the effective width of a filter which is 
\begin{equation}
    \Delta^{'}(\lambda) = \frac{\Delta}{\mathcal{T}(\lambda)}\frac{\lambda_0}{\lambda} 
\end{equation}
where $\lambda_0$ is the mean wavelength of the filter and $\Delta$ is the width of the filter, namely,
\begin{equation}
    \begin{split}
        \lambda_0 & = \frac{\int \lambda\mathcal{T}(\lambda)d\lambda}{\int \mathcal{T}(\lambda)d\lambda}\\
        \Delta & = \int \mathcal{T}(\lambda)d\lambda
    \end{split}
\end{equation}

The remaining key unknown to get the H$\alpha$ of Ion1 is the line ratios of H$\alpha$ and [NII] doublet. Because of the lack of any strong evidence of the AGN presence in Ion1 (see main text), we assume the typical value of star-forming galaxies, i.e. $I_{\rm{[NII]6584}}/I_{\rm{H}\alpha}$$=$0.32 that was initially reported by \citet{Kennicutt1992} and has been widely assumed in the literature. As for the ratio of the two [NII] lines, we assume it to be  $I_{\rm{[NII]6584}}/I_{\rm{[NII]6548}}$$=$3 following \citet{Osterbrock1989}. With these, the equation \ref{equ:mean_eff_wid} now becomes to 
\begin{equation}
    \Delta^{''}(\lambda_{\rm{obs}}^{\rm{H}\alpha}) = \left[\frac{1}{\Delta^{'}(\lambda_{\rm{obs}}^{\rm{H}\alpha})}+0.32\cdot\left( \frac{1}{3\cdot\Delta^{'}(\lambda_{\rm{obs}}^{\rm{[NII]6548}})}+\frac{1}{\Delta^{'}(\lambda_{\rm{obs}}^{\rm{[NII]6584}})} \right) \right]^{-1}.
\end{equation}

Finally, we get the total H$\alpha$ intensity of Ion1 following
\begin{equation}
    \langle f_{\lambda}^{\rm{F356W}}\rangle - \langle f_{\lambda}^{\rm{F335M}}\rangle = \langle f_{\rm{c}}^{\rm{F356W}}\rangle - \langle f_{\rm{c}}^{\rm{F335M}}\rangle + I_{\rm{H}\alpha}\cdot\left[\frac{1}{\Delta^{''}(\lambda_{\rm{obs}}^{\rm{H}\alpha})}\Bigg\vert_{\rm{F356W}}-\frac{1}{\Delta^{''}(\lambda_{\rm{obs}}^{\rm{H}\alpha})}\Bigg\vert_{\rm{F335M}} \right]
\end{equation}
where Ion1's $\langle f_{\rm{c}}\rangle$ in the two filters are estimated using the best-fit SED of Ion1 after masking out H$\alpha$ and [NII] doublet emission. The H$\alpha$ intensity map is obtained by subtracting the PSF-matched F335M image from the F356W image.

\end{document}